\documentclass[useAMS,usenatbib]{mnras}

\usepackage{graphicx}   
\usepackage{amsmath}    
\usepackage{amssymb}    
\usepackage{multicol}   
\usepackage{bm}         
\usepackage{pdflscape}  
\usepackage{float}      
\usepackage{tipa}       
\usepackage[export]{adjustbox}
\usepackage{orcidlink}
\usepackage{fix-cm}
\usepackage{makecell}

\hypersetup{final}

\newcommand{\cms}{\rm \ cm^{-2}}

\title[]{WASP-12, shrouded in mystery or just cold gas?}

\author[Simon Daley-Yates, Ricarda S. Beckmann, Lewis McCallum, Moira M. Jardine, Andrew C. Cameron]{Simon Daley-Yates$^{1,2}$\thanks{E-mail: sddy1@st-andrews.ac.uk} \orcidlink{0000-0002-0461-3029},
    Ricarda S. Beckmann$^{3}$ \orcidlink{0000-0002-2850-0192},
    Lewis McCallum$^{1,4}$ \orcidlink{0009-0007-1181-8034}, \newauthor
    Moira M. Jardine$^{1}$ \orcidlink{0000-0002-1466-5236}, 
    Andrew C. Cameron$^{1}$ \orcidlink{0000-0002-8863-7828}\\
    $^{1}$School of Physics and Astronomy, University of St Andrews, North Haugh, St Andrews, Fife, Scotland KY16 YSS, UK\\
    $^{2}$School of Mathematics and Statistics, University of St Andrews, North Haugh, St Andrews, Fife, Scotland KY16 YSS, UK\\
    $^{3}$Institute for Astronomy, University of Edinburgh, Royal Observatory, Edinburgh EH9 3HJ, UK\\
    $^{4}$RWTH Aachen University, Sommerfeldstr. 16, 52074 Aachen, Germany\\
}

\begin{document}

\date{}

\pagerange{\pageref{firstpage}--\pageref{lastpage}} \pubyear{2023}

\maketitle

\label{firstpage}

\begin{abstract}
Observations of the planet-hosting star WASP-12 show a distinctive depression in the \ion{Mg}{ii} and \ion{Ca}{ii} resonance lines. This has been interpreted as a marker of atmospheric loss from the close-in hot Jupiter WASP-12b and the resulting formation of a gas torus around the star. In this paper we quantify the \ion{Mg}{ii} absorption from this torus, compared to that provided by the stellar wind, the stellar astrosphere and the ISM. To do this we piece together the full density profile of \ion{Mg}{ii} from WASP-12 to an observer on Earth using a combination of hydrodynamical simulations and observations. We find that the bulk of the gas along the line of sight is contained within a dense torus close to WASP-12. However, the temperatures in this torus are sufficient to promote Mg into a doubly (\ion{Mg}{iii}) or higher ionized state. As a result, the singly ionized fraction (\ion{Mg}{ii}) is low. We find that most of the \ion{Mg}{ii} is not in the torus but in the ISM. Despite this, the total column density of \ion{Mg}{ii} is two orders of magnitude lower than required to explain observations of the system. To resolve this discrepancy, we note that the torus gas is at a temperature where it will cool efficiently. We speculate that the onset of the cooling instability will cause the torus to fragment, forming cold clumps with a higher fraction of \ion{Mg}{ii}, capable of explaining the observed absorption.
\end{abstract}


\section{Introduction}

The star WASP-12 is well known to show an unusual spectral feature: there are significant depressions in the chromospheric emission cores of the \ion{Mg}{ii} \citep{Haswell2012} and \ion{Ca}{ii} \citep{Fossati2013} lines. These lines are typically very prominent for F-type stars like WASP-12. The absence of these resonance lines is unexpected for the spectral type and age of the star \citep{Fossati2010}. While it was first suggested that this lack of emission lines might be intrinsic, the required low stellar activity has been ruled out \citep{Fossati2013,Bonomo2017}. 

Alternatively, the emission lines could be absorbed by material along the line of sight (LOS). \citet{Haswell2012} have estimated that a \ion{Mg}{ii} column density of at least $2 \times 10^{17} \cms$ is required to convert the spectrum of a typical F-type star to those observed from WASP-12. \citet{Haswell2012} combine a Mg+ density estimate from standard Milky Way reddening laws with an estimate of the ionization fraction of \ion{Mg}{ii} of 0.5 to estimate that the maximum contribution of the interstellar medium (ISM) to the \ion{Mg}{ii} column density is $2 \times 10^{16} \cms$. The ionization fraction was estimated using observations of Alpha Centauri, which lies along the same LOS as WASP-12 but is much closer. This estimate of the \ion{Mg}{ii} ISM column density, together with the requirement of a chance alignment of the stellar radial velocity with that of the intervening ISM along the LOS, shows that it is unlikely that the unique features of WASP-12 are due to ISM absorption.

This means that there must be a dense feature along the LOS, most likely near the star. As well as the central F-type star, the WASP-12 system contains WASP-12 b, one of the most extreme hot Jupiters known \citet{hebb2009WASP12bHottestTransiting}. With an orbital radius of only 0.0232au (1 stellar diameter from the surface), and an orbital period of just 1.1 days, WASP-12 b is one of the most irradiated planets \citep{hebb2009WASP12bHottestTransiting}. Near-UV transit spectroscopy by \citep{Haswell2012} with the Cosmic Origins Spectrograph \citep{Osterman2011} aboard HST shows excess transit depths in strong lines, indicating extensive diffuse gas around WASP-12 b, extending significantly beyond the Roche lobe \citep{Haswell2012}. If the gas is dense enough it could be responsible for the absorption along the LOS.

A range of sources has been suggested as the origin of the dense gas. The outer planetary layers could be removed from Wasp-12 b by the intense radiation \citep{vidal-madjar2004DetectionOxygenCarbon,guo2011EscapingParticleFluxes,ehrenreich2011MasslossRatesTransiting} or by tidal interactions \citep{li2010WASP12bProlateInflated}. The latter theory may be less likely since it was shown that WASP-12 has very low eccentricity \citep{lopez-morales2010DaysideBandEmission,husnoo2011OrbitalEccentricityWASP12}. However, a secular decrease in the orbital period, observed by \cite{Leonardi2024}, likely attributable to tidal orbit decay, is now well established through long-term transit timing observations. This suggests that WASP-12b is undergoing fast tidal dissipation and adds credence to the idea that tidal interaction is responsible for the dense gas. Other suggestions for the source of the material included Roche lobe overflow \citep{Lai2010} or entrained material from the stellar corona \citep{vidotto2010EarlyUVIngress,llama2011ShockingTransitWASP12b}.

 \cite{Debrecht2018} simulated the early evolution of a model in which a planetary wind, driven by the radiation from the star, interacts with the stellar wind resulting in a torus around WASP-12. The gas density in the torus, fed by the combined winds, increases steadily over time at a rate of $8.09 \times 10^{-18} \mathrm{g}/\mathrm{cm}$ per orbit. Extrapolating from their simulation, they estimate that it would take approximately 13 years for the torus to reach the required density of $3.65 \times 10^{-14} \mathrm{g}/\mathrm{cm}$ required to produce the \ion{Mg}{ii} column density of $2 \times 10^{17} \mathrm{cm}$ that is observed in the resonance line absorption. However, they were unable to demonstrate that the required densities are actually reached in the torus as their simulation only covered 14 orbits ($\sim$15.4 days), which leaves uncertainties about the final steady-state configuration. Their work demonstrates that planetary mass loss can form a translucent circumstellar structure potentially responsible for the observed \ion{Mg}{ii} and Ca II h\&k line anomalies, while highlighting the need for extended simulations to confirm the torus' ultimate stability and density.

In this paper, we use numerical simulations to investigate whether the cumulative effect of absorption in the dense torus and the intervening ISM is sufficient to explain the observed peculiar spectral features of the WASP-12 system. Our aim for this study is to join up a complete profile of the density structure of the space between Earth and the WASP-12 system. The paper is structured as follows: We present the simulations analysed in this paper in Section \ref{sec:modelling} including the torus close to the star (Section \ref{sec:method_torus}), the astrosphere (Section \ref{sec:astrosphere_sim}) and the model for interstellar medium in Section \ref{sec:ism_profile_modelling}. Results are discussed in Section \ref{sec:results} and conclusions presented in Section \ref{sec:conclusions}.

\begin{table*}
        \centering
        \caption[]{Simulation parameters used in both the torus and Astrosphere simulations. {$\ast$ refers to the star while $\circ$ refers to the planet}.  \label{tab:parameters}}
        \begin{tabular}{cccccc}
                \hline 
                Parameter & Symbol & Star & Planet & Comment \\
                \hline
                Mass & $M_{\ast, \circ}$ & $1.434 \ M_{\odot}$ & $1.47 \ M_{J}$ & from \cite{Chakrabarty2019} \\
                Radius & $R_{\ast, \circ}$ & $1.657 \ R_{\odot}$ & $1.9 \ R_{J}$ & from \cite{Chakrabarty2019} \\
                {Wind} temperature & $T_{\ast,\circ}$ & $1.6\times10^{6} \ \mathrm{K}$ & $5 \times 10^{4} \ \mathrm{K}$ & from \cite{Debrecht2018} \\
                {Mass loss rate} & {$\dot{m}_{\ast,\circ}$} & {$6 \times 10^9 \mathrm{g/s}$} & {$10^{12} \mathrm{g/s}$} & \makecell{{From \citet{Debrecht2018} and} \\ {\citet{ehrenreich2011MasslossRatesTransiting} respectively}} \\ 
                {Wind} density & $\rho_{\ast, \circ}$ & $5 \times 10^{-15} \ \mathrm{g}/\mathrm{cm}^{3}$ & $2 \times 10^{-14} \ \mathrm{g}/\mathrm{cm}^{3}$ & Parametrised for required mass loss rate \\
                Orbital radius & $a$ & $-$ & 0.0234 au & from \cite{Ozturk2019} \\
                {WASP-12 radial velocity} & {$v_{\mathrm{RV}}$} & {19.46 $\mathrm{km}/\mathrm{s}$} & - & {From \cite{Czesla2024}} & \\
                WASP-12 frame ISM velocity & $v_{\mathrm{ISM}}$ & 23.72 $\mathrm{km}/\mathrm{s}$ & - & \makecell{{Derived from \cite{Czesla2024} and} \\ {\cite{Gaia32023}}} & \\
                ISM mass density & $\rho_{\mathrm{ISM}}$ & 2.22$\times 10^{-25} \mathrm{g}/\mathrm{cm}^{3}$ & - & From ISM profile, see Section \ref{sec:ism_profile_modelling} & \\
                \hline
        \end{tabular}
\end{table*}

\section{Modelling}
\label{sec:modelling}
To piece together the full column density profile from WASP-12 to an observer on Earth, we combine a series of three simulations:
\begin{itemize}
    \item[1.] A 3D hydrodynamic simulation of the inner system including both the star and planet, which extends from the surface of the star $R_\ast$ to $10 R_{\ast}$ and is designed to capture the build up of planetary gas in the system. We will refer to this simulation as the torus simulation from here on and present its details in Section \ref{sec:method_torus}.
    \item[2.] A hydrodynamic simulation that covers the region 100 - $5 \times 10^{5}$au from WASP-12 {($\sim 1.2 \times 10^4 - 6 \times 10^7 \rm \ R_{\ast}$)} and is designed to capture the interface between the stellar wind and the inter-stellar medium (ISM). This simulation will be referred to as the Astrosphere simulation. Its details are presented in Section \ref{sec:astrosphere_sim}.
    \item[3.] A 1D LOS profile of the density and ionization from WASP-12 to the solar system, taken from a 3D simulation. We present the details of this method in Section \ref{sec:ism_profile_modelling}. 
\end{itemize}
We explain how we combine information from all three simulations to predict the full density profile from WASP-12 to an observer on Earth in Section \ref{sec:total_column_density}.

\subsection{Hydrodynamics}
For both the torus and the astrosphere simulation, we solve the following system of hydrodynamic equations using the static grid version of the public code PLUTO (version 4.3) \citep{Mignone2007}.
\begin{equation}
\partial_{t}
\begin{pmatrix}
\rho \\
\bm{m} \\ 
E
\end{pmatrix}
+ \bm{\nabla} \cdot
\begin{pmatrix} 
\rho {\bm{v}} \\ 
\bm{m} \bm{v} + p \bm{I} \\
\bm{v} \left( E + p \right)
\end{pmatrix}
=
\begin{pmatrix} 0  \\ 
\rho \pmb{\textsl{g}} \\ 
\bm{m} \pmb{\textsl{g}}
\end{pmatrix}
\label{eq:Euler_seperate_full}
\end{equation}
Here $E$ is the total energy {density} according to:
\begin{equation}
    E = \frac{p}{\gamma - 1} + \frac{|\bm{m}|^{2}}{2\rho},
\end{equation}
$p$ is the thermal pressure, $\bm{m}$ is the momentum vector, $\rho$ is the density, $\bm{v}$ is the velocity vector, $\gamma$ is the ratio of specific heats and $\bm{I}$ is the identity matrix. {We convert between mass density and number density by $\rho=\mu m_{p} n$ where $n$ is the total particle number density and $\mu = 0.6$ is the mean molecular weight of the gas in our simulation.}

Both the torus and Astrosphere simulations use the same underlying physics of ideal hydrodynamics. There are two major differences in that the torus simulation is done in the rotating frame of WASP-12b's orbit and has $\gamma = 1.05$. This $\gamma$ value mimics the effect of cooling, heating and thermal conduction in the stellar wind close to the star without the need to explicitly include these physics and the numerical limitations they bring. The Astrosphere simulation is done in the {co-moving frame of the system}. A $\gamma=5/3$ is used as the effects of coronal physics are negligible out at the distances of WASP-12's astrosphere.

\subsection{Torus simulation}
\label{sec:method_torus}

In this work we present a 3D simulation of the environment of WASP-12 that extends from the stellar surface to {$10 \ R_*$ (the outer edge of the computational domain)}. We present here the relevant parameters used for simulating the WASP-12 system but refer the reader for greater details of the specific numerical grid and schemes to \cite{DaleyYates2019}.

The torus simulation covers the range $r~\in~\{ 1\ R_{\ast}, 10 \ R_{\ast} \}$, $\theta~\in~\{ 0, \pi \}$ and $\phi~\in~\{ 0, 2 \pi \}$ with a resolution of $284 \times 210 \times 384$ respectively. Unlike in \cite{DaleyYates2019} we are able to extend the computational domain to include the polar singularities as we make use of the polar-type boundary conditions and the ring averaging method of \cite{Binzheng2019} which are implemented in PLUTO by its developers (see the release notes and user guide of version 4.4 of PLUTO, contemporaneous to this paper). The ring averaging technique allowed the simulation to make time-steps approximately seven times larger than would be required without it, which speeds up the computation and allows our simulation to cover the long timescales required to reach convergence. 
 
The simulation was run for 75 orbital periods $t_{\rm orbit}$. As for WASP-12, $t_{\rm orbit} \sim 1.1 \rm \ d$, our simulation therefore covers $ 82.2 \rm \ d$. Boundaries are set to outflow, so material can escape the computational domain. 

Within the simulation we include two boundaries that represent the star and the planet separately: The lower radial boundary located at $R_{\ast} = 1.657 R_{\sun}$ represents the star WASP-12, and a smaller sphere {with a radius $R_\circ = 1.9 \rm R_{\rm J}$} located at a distance of $0.0234 \rm au$ ($\sim 3 R_{\ast}$)  along the x-axis, represents the planet WASP-12b (See Tab. \ref{tab:parameters} for model parameters). The surface of both the star and planet are modelled to drive a Parker wind \citep{Parker1965} { The winds are assumed to have a temperature of $1.6 \times 10^6 \ K$ for the star and $5 \times 10^4 K$   for the planet, following \citet{Debrecht2018}. We compute the wind velocity at injection following the Parker model, and set the wind density to ensure a constant mass-loss rate of $6\times10^{9} \ \rm g s^{-1}$ for the star and $10^{12}  \ \rm g  s^{-1}$ \citep{ehrenreich2011MasslossRatesTransiting} for the planet respectively.} The circumstellar medium around the star and planet is initialised using a Parker wind solution based on the wind from the star. The properties of the star {(subscript $\star$)} and planet {(subscript $\circ$)} are summarised in table \ref{tab:parameters}. 

{When calculating LOS column densities from this simulation we use an inclination  $i_{\mathrm{obs}}=84.955^{\mathrm{o}} \pm 0.037$ obtained by \cite{Turner2021} who achieved this precision by modelling the planet’s transit  profile. We see no reason for the torus not be coplanar with the planet’s orbit, therefore we assume the inclination of the torus should be the same as the planets orbit.}

\subsection{Astrosphere simulation}
\label{sec:astrosphere_sim}
We model the astrosphere of WASP-12 using a {a 2D simulation with assumed cylindrical symmetry about the $x$-axis which allows us to orientate the simulation so that the LOS coincides with the plane of the simulation}. Our simulation is set up in a manner roughly similar to \cite{Wood2005}, but we do not treat charge exchange and limit ourselves to ideal hydrodynamics. 

{The 2D polar grid has the following extent $r~\in~\{ 100 \ \mathrm{au}, \ 5\times10^{5} \ \mathrm{au} \}$ and $\theta~\in~\{ 0, \pi \}$ with a resolution of $512 \times 256$ respectively. The grid was logarithmically stretched in the radial direction to keep the cell aspect ratio approximately unity. The numerical scheme used was parabolic spatial reconstruction paired with 3rd order Runge-Kutta time integration and Harten-Lax-van Leer-Contact discontinuity Riemann solver (HLLC). }

We perform the simulation in the co-moving frame of WASP-12. Therefore, the system sees the ISM move past it with a velocity equal to the true velocity of the system. We calculate the true velocity by combining WASP-12's proper motion, $13.57 \rm \ km \ s^{-1}$ (obtained from the Gaia 3 data release \citep{Gaia32023}, RA:~-~1.519~mas~yr$^{-1}$, Dec.:~-~6.761~mas~yr$^{-1}$), and the systemic radial velocity, $19.46 \rm \ km \ s^{-1}$ \citep{Czesla2024}. {The result is a true velocity of WASP-12 relative to the local ISM of $v_{\rm rel,ism} = 23.72 \rm \ km \ s^{-1}$.}

The simulation is initialized with a Parker wind solution, similarly to the torus simulation, to a radius of $r < 3000$au. The inner radial boundary condition is held constant and ensures the Parker wind solution is maintained up to the magnetopause while the simulation is evolving. Beyond $r > 3000$au, the initial conditions are a smooth ISM with density $\rho_{\rm ism} = 2.22 \times 10^{-25} \mathrm{g}/\mathrm{cm}^{3}$ and a velocity $v_{\rm rel,ism}$. {The same is true within the boundary layer.} The choice of $\rho_{\rm ism}$ is based on our ISM profile and explained in Subsection \ref{sec:ism_profile_modelling}. The ISM extends to the upper radial boundary at $r~=~5\times10^{5} \ \mathrm{au}$ and is inflowing where $0~\leq~\theta~\leq~\pi/2$ and outflowing where $\pi/2~\leq~\theta~\leq~\pi$. The $\theta$ boundaries are set to reflective.

\subsection{Interstellar medium profile}
\label{sec:ism_profile_modelling}

To construct the hydrogen density along the LOS from the Sun to WASP-12 we follow the method in \citet{mccallum2025HaSkyThree} which is based on the 3D mean differential extinction map within 1.25 kpc of the Sun from \citet{edenhofer2024ParsecscaleGalactic3D}, which contains the location of WASP-12 at 413 pc from Earth \citep{Gaia32023}. The dust map is converted to total hydrogen density following \citet{oneill2024LocalBubbleLocal}, based on work by \citet{zucker2021ThreedimensionalStructureLocal}, which combines the extinction curves from \citet{zhang2023Parameters220Million} with the extinction-to-column density factor from \citet{draine2009InterstellarDustModels} (see \citet{mccallum2025HaSkyThree} for details). The resulting density is interpolated onto a regular $1024^3$ Cartesian grid using the interpolation script shipped with the data from \citet{edenhofer2024ParsecscaleGalactic3D}. To estimate the fraction of \ion{Mg}{ii} we assume a Mg/H abundance of $3.5\times 10^{-5}$, as in the rest of the paper \citep{suess1956AbundancesElements}. We then calculate the ionization state of the gas as in \citet{mccallum2025HaSkyThree}, by carrying out a static photoionization simulation using a catalogue of nearby O-stars. This enables the calculation of the ionization fraction of various astrophysical ions, including \ion{Mg}{ii}. We then extract the 1D profile between the Sun and WASP-12. 

\section{Results}
\label{sec:results}

\subsection{Density features along the line of sight}
In this section we piece together the mass density profile along the LOS from WASP-12 over the entire 413 pc from the stellar surface to the solar neighbourhood. We work from WASP-12 outwards.

\subsubsection{The torus: WASP-12 system gas structure}
\label{sec:torus_profile}

\begin{figure*}
	\centering
	\includegraphics[width=0.99\textwidth,trim={0cm, 0cm, 0cm, 0cm},clip]{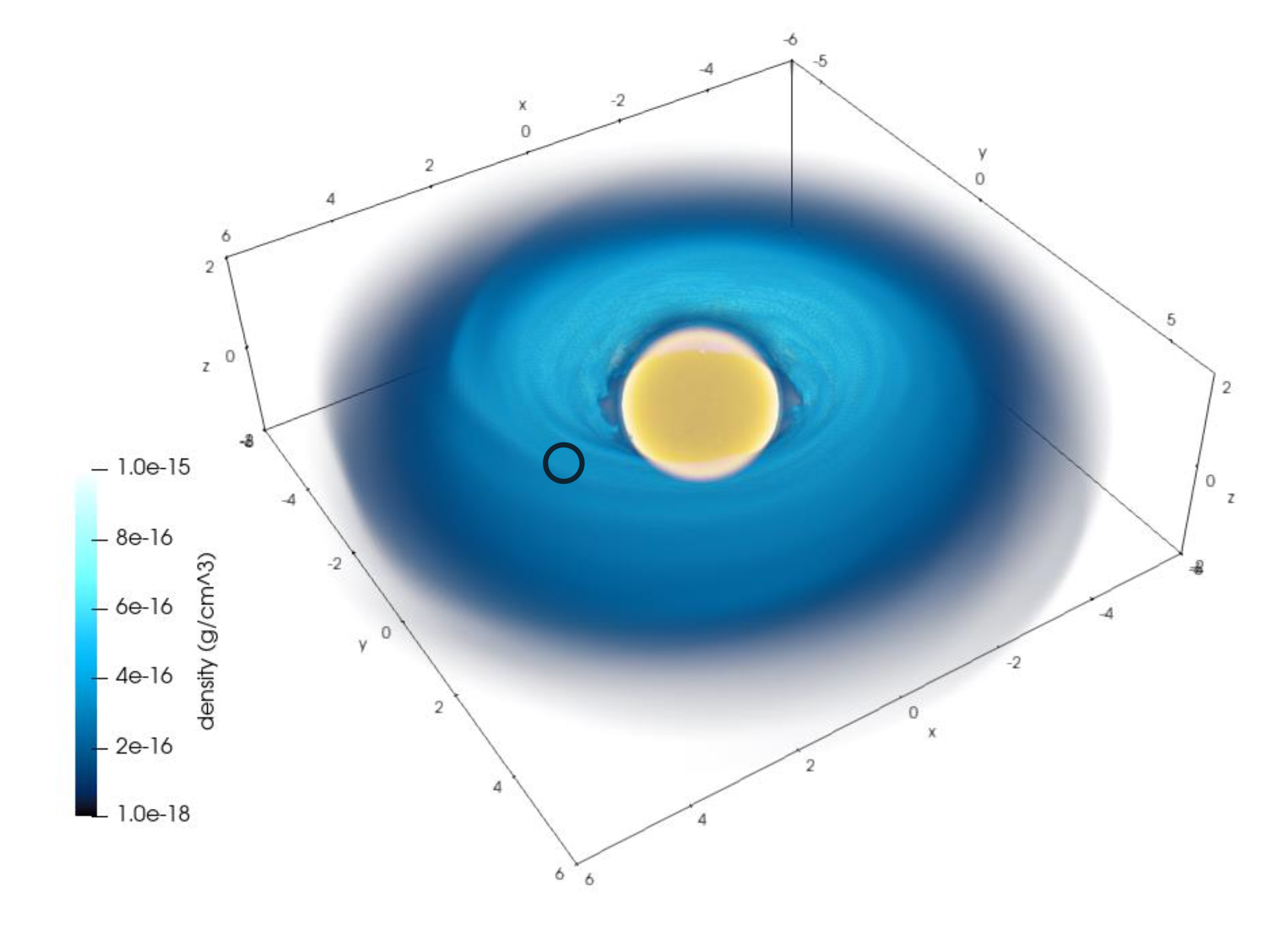}
	\caption[]{{Volume rendering of the 3D gas density distribution (blue colour map) around WASP-12 after 60 orbits. The star is located in the centre (yellow), with the planet embedded in the torus, offset along the $x$-axis from the star and is highlighted by the black circle.} \label{fig:volume}}
\end{figure*}

\begin{figure*}
	\centering
	\includegraphics[width=0.99\textwidth,trim={0cm, 0cm, 0cm, 0cm},clip,left]{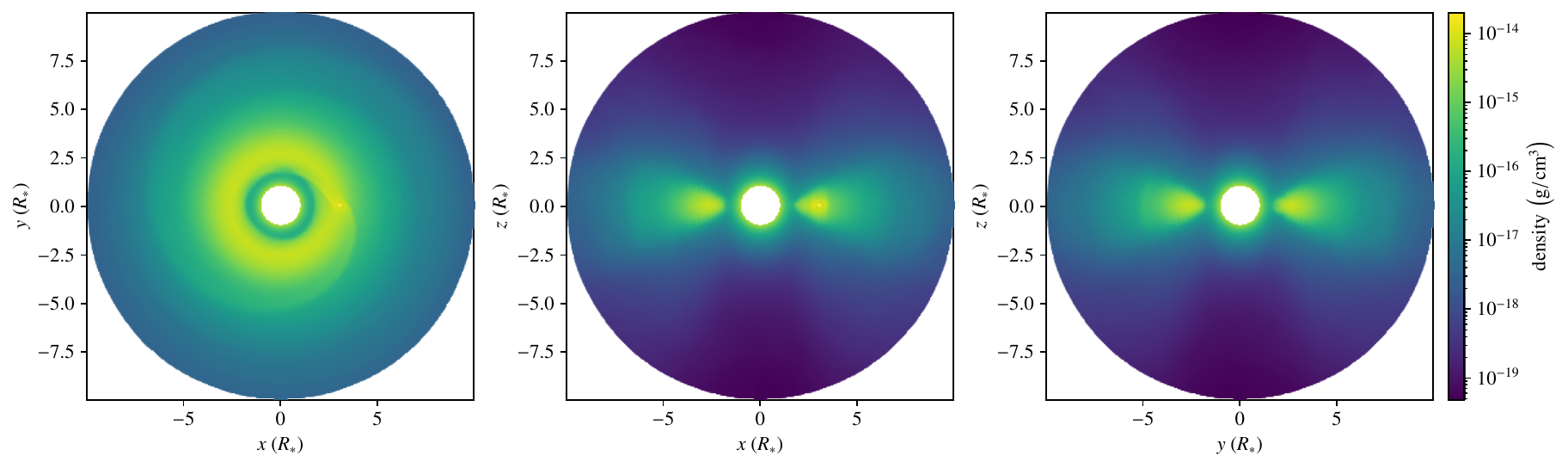}
	\includegraphics[width=0.9725\textwidth,trim={0cm, 0cm, 0cm, 0cm},clip,left]{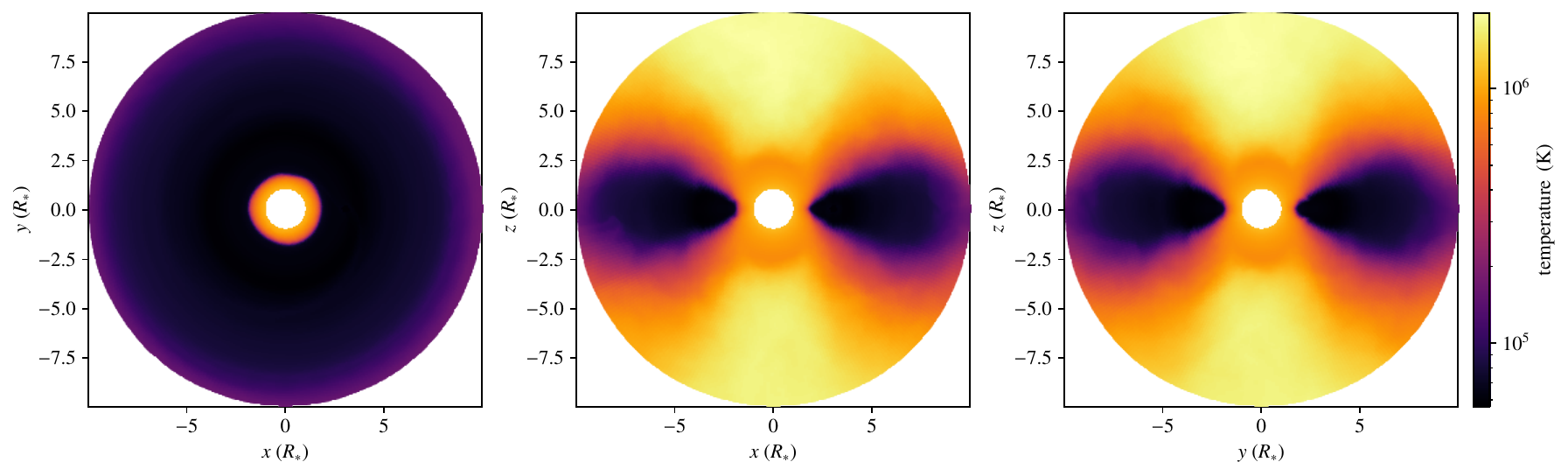}
	\includegraphics[width=0.9725\textwidth,trim={0cm, 0cm, 0cm, 0cm},clip,left]{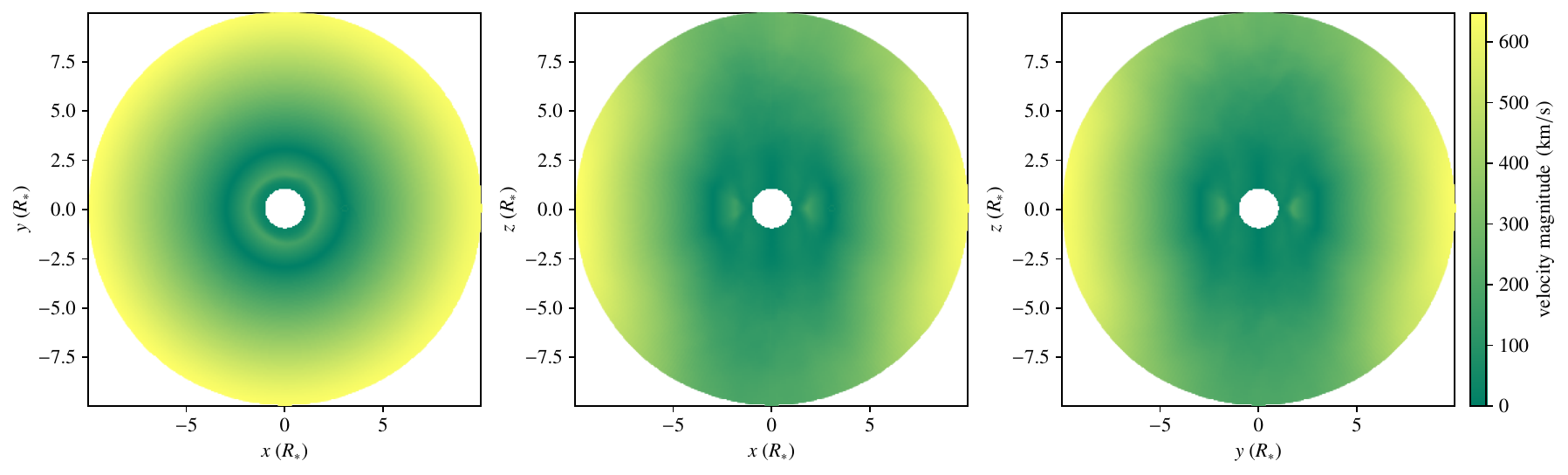}
	\includegraphics[width=0.9725\textwidth,trim={0cm, 0cm, 0cm, 0cm},clip,left]{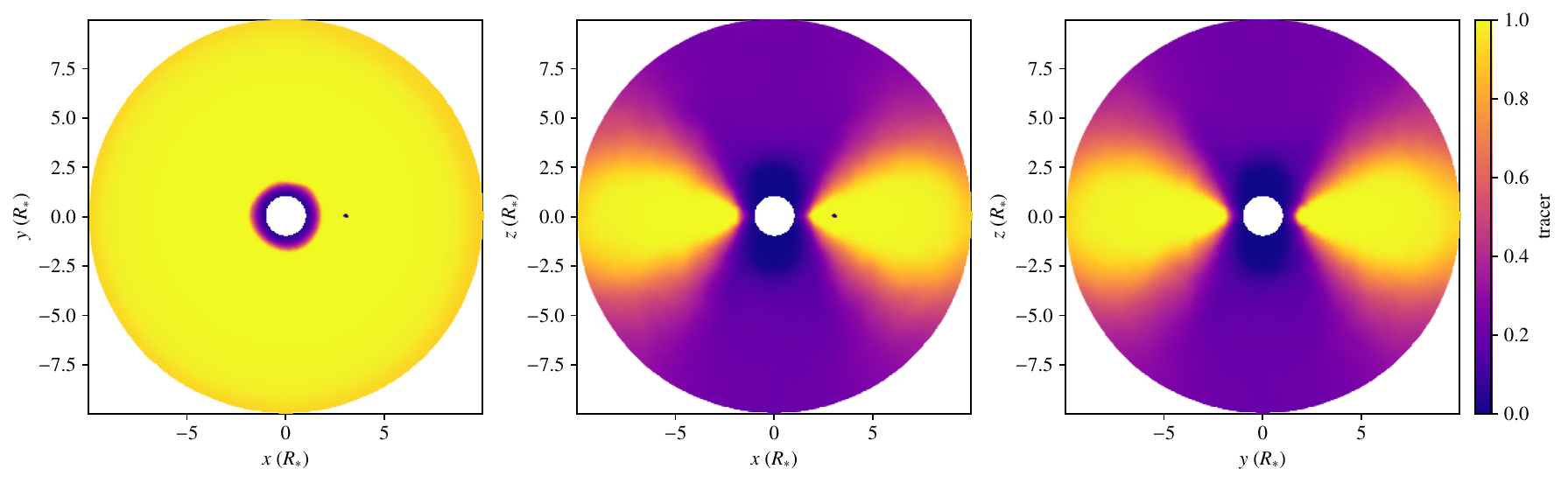}
	\caption[]{Slice plots time-averaged over the last 10 orbits of the planet for three different views: top down (left), side on in the plane of the planet's position (middle) and side on perpendicular to the plane of the planet's position. Time averaging is done in the rotating frame of the planet, which is held fixed at orbital separation for the plots shown here. For this reason the downstream spiral structure behind the planet remains visible despite the time averaging. 
 From top to bottom: density, temperature, velocity magnitude and tracer concentration. \label{fig:profiles}}
\end{figure*}

\begin{figure}
	\centering
	\includegraphics[width=0.49\textwidth,trim={0cm, 0cm, 0cm, 0cm},clip]{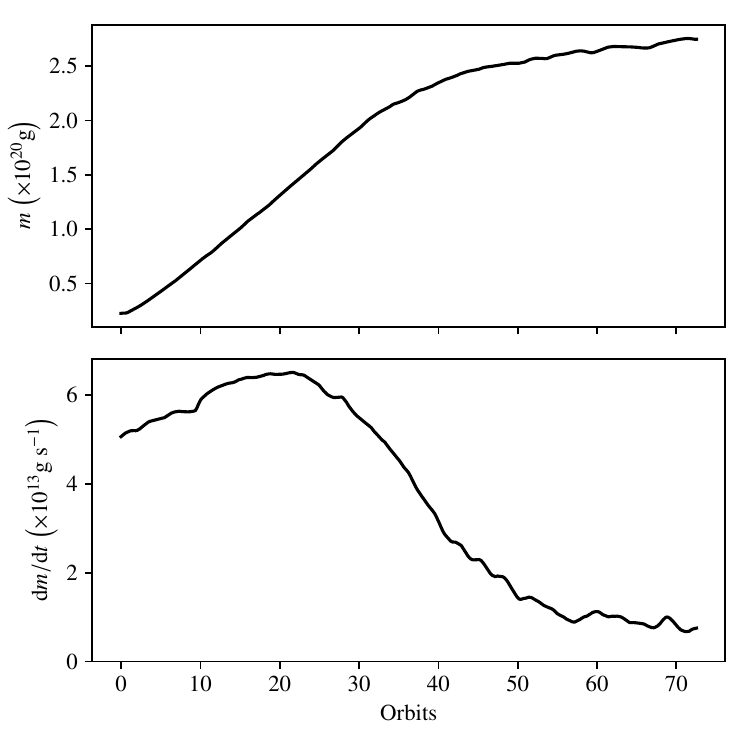}
	\caption[]{Top: time series of (hydrogen) gas mass-build-up in the simulation domain. At late times the curve has flattened off, indicating convergence of the simulation. Bottom: block average (over 64 outputs at a time) of the rate of change of total mass in the simulation. {This method smooths out the stochastic fluctuations in the total mass, ensuring $\mathrm{d}m/\mathrm{d}t > 0$ at all times.} We use this as an indicator of the convergence of the simulation. While the value of $\mathrm{d}m/\mathrm{d}t$ has not reached zero by $t=75 t_{\rm orbit}$, we deem that the rate of mass-build-up has slowed sufficiently to consider the simulation converged. \label{fig:mass_time}}
\end{figure}

\begin{figure}
	\centering
	\includegraphics[width=0.49\textwidth,trim={0cm, 0cm, 0cm, 0cm},clip]{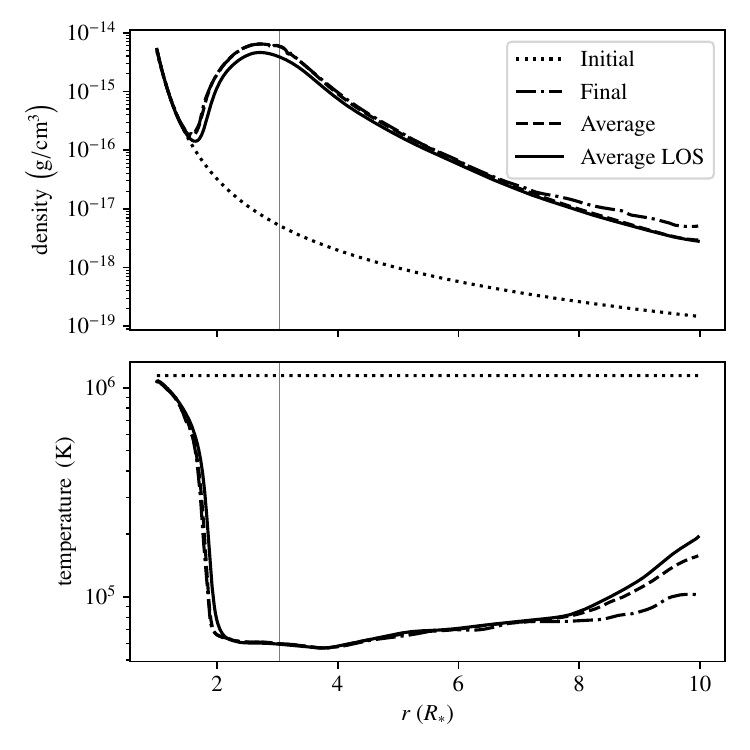}
	\caption[]{Radial mass density (top) and temperature (bottom) profile at times $t=0$ (dotted, the initial, pure Parker wind from the star) and $t=75 \rm \ t_{\rm orbit}$ (dot-dashed) through the orbital plane of the planet. Also shown are profiles averaged over the last 10 orbits, both along the LOS (solid) and through the plane of the planet (dashed). The location of the planet is marked by the vertical line. {Profiles shown here are computed at an azimuthal angle of 180 degrees from the planet.} \label{fig:torus_density} }
\end{figure}

First we present the evolution of the torus simulation which dominates the gas density distribution within a few stellar radii $R_*$ of WASP-12. As can be seen in Fig. \ref{fig:volume}, over time a dense gas torus builds up around the central star. Gas ejected from the planet by the planetary wind slowly fills the planet's orbit. The torus is denser in the middle and becomes more diffuse both towards the star and outwards where it diffuses out in a spiral structure at a few stellar radii. The inner edge, close to the star, is unstable where the diffuse wind from the central star impacts the denser torus. The gas falls under the gravity of the central star in a similar manner to the Rayleigh-Taylor instability and impacts the star's surface.

The time-averaged plots in Fig. \ref{fig:profiles} show the gas structure in the rotating frame of the planet in more detail. The inner edge of the torus is clearly defined. The spiral structure in the gas remains a persistent feature, with both an upstream and a downstream component clearly visible. The edge-on density plots (middle and right column) show that the torus narrows towards its inner edge and flares outwards to the edge of the computational domain. The torus reaches its maximum density coincident with the orbital position of the planet and is surrounded by a more diffuse halo of gas. The temperature plots in Fig. \ref{fig:profiles} show that the central dense torus is colder, at the temperature of the planetary wind ($\sim 10 000 \rm \ K $), while the diffuse halo is hotter, roughly matching the temperature of the stellar wind. We confirmed that the gas in the torus originates from the planetary wind using a hydrodynamic tracer (bottom panels in Fig. \ref{fig:profiles}).

{The planetary wind is injected isotropically from the planetary surface (or more specifically the inner computational domain that represents the planet, see \cite{DaleyYates2019} for more details). As a result the material has an effective distribution of momentum, some towards the star, others along the planets orbit path and some out into the stellar wind. The material then travels under the influence of the joint gravitational potential of the star and planet, and interacts with the stellar wind through ram-pressure effects. This is a non-linear process, as can be seen by the fact that the density of the torus saturates over time, which is why we have chosen to conduct the simulations presented here rather than attempt to derive the density profile of the torus analytically.}

As can be seen in Fig. \ref{fig:mass_time}, the mass growth of the torus can be split into roughly two regimes: an early linear growth phase at time $t \lesssim 25 \ t_{\rm orbit}$, and a late saturation phase at $ t > t_{\rm orbit}$. In the early phase, the mass in the torus grows linearly. As can be seen in the bottom panel of Fig. \ref{fig:mass_time}, the mass growth of the torus during the early phase is not strictly linear, as the mass growth rate slowly increases from $5 \times 10^{13} \rm \ g \ s^{-1}$ at the beginning of the simulation to a maximum of $6.2 \times 10^{13} \rm \ g \ s^{-1}$ at $t \approx 25\ t_{\rm orbit}$.The linear growth phase ends around 25 orbits when the torus mass starts to saturate. While the torus continues to grow in the time interval $25 - 60 \ t_{\rm orbit} $ the growth rate continues to drop during this time. We consider the simulation converged from $t=60 \ t_{\rm orbit}$ onwards, when the growth rate has fallen to $< 10^{13} \rm \ g \ s^{-1}$.
 At this point, the continued mass input from the planetary wind is closely balanced by the mass outflow at the outer edge of the torus. The time-averaged slice plots shown in Fig. \ref{fig:profiles} are computed by averaging over all simulation outputs during the last 10 $t_{\rm{orbit}}$.

The early linear growth phase reported here is in agreement with results by \citet{Debrecht2018} who reported approximately linear mass growth of their torus until the end of their simulation time around $t \approx 15 t_{\rm{orbit}}$. They estimated that if the linear growth phase could be maintained for another 13 years, the column density would become sufficiently high to explain the observed absorption. Unfortunately, our simulations show that such sustained growth is extremely unlikely, as the growth rate slows down after just $\sim 35$ days, and total mass in the torus saturates after about 65 days ($60 \ t_{\rm orbit}$). We note that all values quoted in this section concern the total mass contained within the simulation box as it is difficult to clearly separate the diffuse torus from the remaining gas. 

The radial density profile of the converged torus is shown in Fig. \ref{fig:torus_density}. The inner edge of the torus is located around $r = 2 R_*$, as can be seen by both the increase in density and the decrease in temperature. Within the central cavity, the density profile remains dominated by the stellar wind. Perturbations of the instantaneous (final) profile are small in comparison to the average. For the rest of this section, we will discuss the time-averaged profile.

The peak density of $6.29\times 10^{-15} \ \rm{g}\ \rm{cm}^{-3}$ is located at $1.814~\ R_{\ast}$ (0.0140au) from the stellar surface, {which is closer to the star than the planet which is located at 0.0234au}. From the peak, the density falls off approximately exponentially and remains enhanced to the edge of the simulation domain in comparison to the density expected from the stellar wind alone (dotted line). As the LOS does not pass through the orbital plane of the planet, it also misses the densest region of the torus, which can be found in the mid-plane. 

{This results in a peak LOS density (solid) of $4.5\times 10^{-15} \ \rm{g}\ \rm{cm}^{-3}$, a factor of 1.4 lower than the peak density in the mid-plane (dashed) but the two converge at large distances from the star. As the plots in the left-hand column of left-hand  Fig. \ref{fig:profiles} show, the azimuthal variations of 
density around the torus are small, except in the very narrow region around the planet 
where somewhat higher densities are found. We have chosen to compute the density 
profiles shown in Fig. 4 at 180 degrees from the planet. If we consider all possible values, 
the variations in density along the line of sight are not large, ranging between 0.97 to 1.2 
times the LOS density shown in Fig. \ref{fig:torus_density}. This means that the expected modulation of the absorption over the orbital frequency ($\sim$1 day) is small.}

The temperature profile in the lower panel of Fig. \ref{fig:torus_density} confirms the trend with temperature already evident in the temperature maps in Fig. \ref{fig:profiles}: the dense torus is significantly colder than the stellar wind, with a temperature profile closer to the $10^4 \rm \ K$ of the planetary wind than the $10^5 \rm \ K$ of the stellar wind. Towards the edge of the simulation domain, as the torus becomes more diffuse, the temperature increases towards that of the stellar wind. From the trends in both the temperature and density profiles shown here, we expect the solution {to have reliably converged back to a Parker wind profile by 1 au  at the latest}(see Section \ref{sec:total_column_density} for further details).

From the density profile in Fig. \ref{fig:torus_density} we compute an integrated column density {along the line of sight} for comparison to observations: after the torus is established, the column mass density reaches $8.112 \times 10^{-4} \ \rm{g}\ \rm{cm}^{-2}$. In comparison to the initial column mass density from a pure stellar wind of $7.551 \times 10^{-5} \ \rm{g}\ \cms $, this is approximately ten times larger. We discuss the potential impact of this gas on the \ion{Mg}{ii} absorption further in Section \ref{sec:total_column_density}.

\subsubsection{The astrosphere}
\label{sec:astrosphere_profile}

\begin{figure*}
	\centering
	\includegraphics[width=0.99\textwidth,trim={0cm, 0cm, 0cm, 0cm},clip]{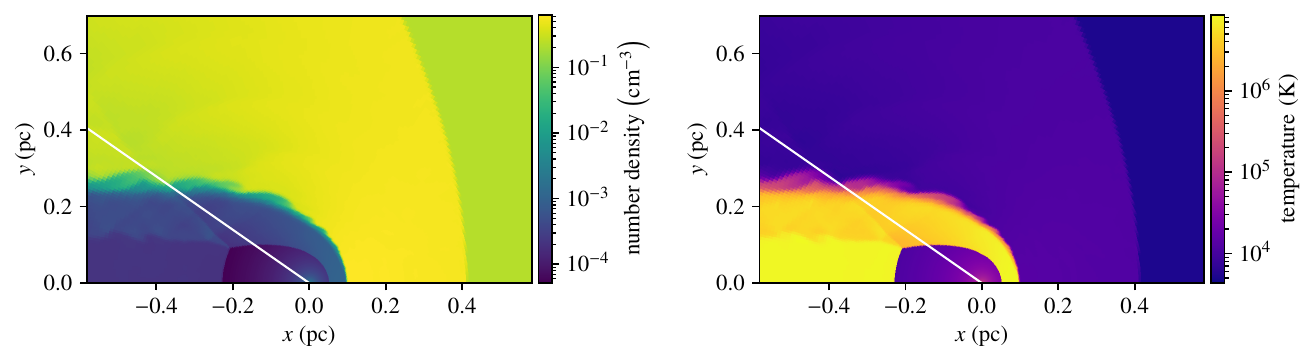}
	\caption[ ]{Number density and temperature of the astrosphere of WASP-12. The system is located at the origin. The simulation is conducted in the co-moving frame with the system, as such the ISM inflows from the right had side of the plot at a velocity of 23.72 km s$^{-1}$ and density of $2.212\times10^{-1} \ \mathrm{cm^{-3}}$ ($2.22 \times 10^{-25} \ \mathrm{g} \ \mathrm{cm^{-3}}$) (see text in Section \ref{sec:astrosphere_sim} for details of these values). This inflow collides with the spherically expanding stellar wind resulting in the formation of a termination shock, and astropause. The structure is assumed to be cyclindrically symmetric about the x-axis. This allows us to conduct the simulation in 2D. The LOS to the observer is annotated by the white line. \label{fig:density_asph}}
\end{figure*}

\begin{figure}
	\centering
	\includegraphics[width=0.49\textwidth,trim={0cm, 0cm, 0cm, 0cm},clip]{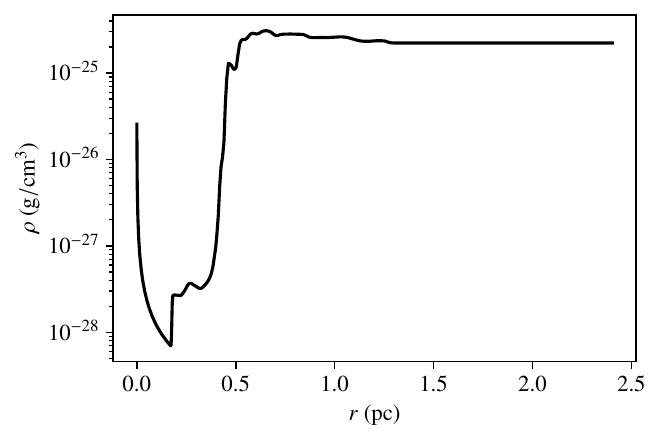}
	\caption[]{Hydrogen density profile of the astrosphere simulation shown in Fig. \ref{fig:density_asph}, covering a distance of $10^2-5\times 10^{5} \rm au$ from WASP-12 along the LOS to the observer. \label{fig:density_asph_profile}}
\end{figure}

The torus simulation covers the region out to $10 \rm \ R_*$ ($\sim 0.077 \rm au$) so we now turn our attention to the evolution of the density along the LOS further away from the star. Stars with stellar winds such as WASP-12 are known to inflate low-density bubbles in the ISM. These so-called astrospheres are edged by dense shocks where the stellar wind impacts and sweeps up the ISM. To understand how the density structure of the astrosphere might contribute to the \ion{Mg}{ii} absorption we conducted a second simulation whose parameters are presented in Section \ref{sec:astrosphere_sim} which bridges the gap between the torus simulation and ISM data ($10^2-5\times 10^{5} \rm au$).

As shown in the density map in Fig. \ref{fig:density_asph}, the combination of the stellar wind and the relative motion of WASP-12 with respect to the ISM produces a cone-shaped under-dense astrosphere with a dense bow-shock. The LOS to WASP-12 passes through the overdensity created by the bow-shock but misses the densest part of the shock ahead of the star. The resulting density profile along the LOS is shown in Fig. \ref{fig:density_asph_profile}. The inner structure of the profile is dominated by the stellar wind. Where the wind impacts the ISM, a shock forms and the density sharply increases before falling back to a value similar to the local ISM mass density of $2.22 \times 10^{-25} \ \mathrm{g} \ \mathrm{cm}^{-3}$. For continuity, we have included a smooth interpolation onto the local ISM density at large distances from WASP-12. 

A comparison of the peak densities in Fig. \ref{fig:density_asph_profile} and the density profile through the torus in Fig. \ref{fig:torus_density} shows that the densest gas along the LOS through the edge of the astrosphere is more than 10 orders of magnitude lower than in the torus. Despite the much larger spatial extent of this region, the astrosphere only contributes a column number density of $1.414\times10^{-6} \ \mathrm{g} \cms$, approximately 3 orders of magnitude less than the torus. We therefore conclude that the structures around the astrosphere are unlikely to significantly contribute to the total column density along the LOS.

\subsubsection{The interstellar medium from WASP-12 to the Sun}
\label{sec:ism_profile}

\begin{figure*}
	\centering
	\includegraphics[width=0.99\textwidth,trim={0cm, 0cm, 0cm, 0cm},clip]{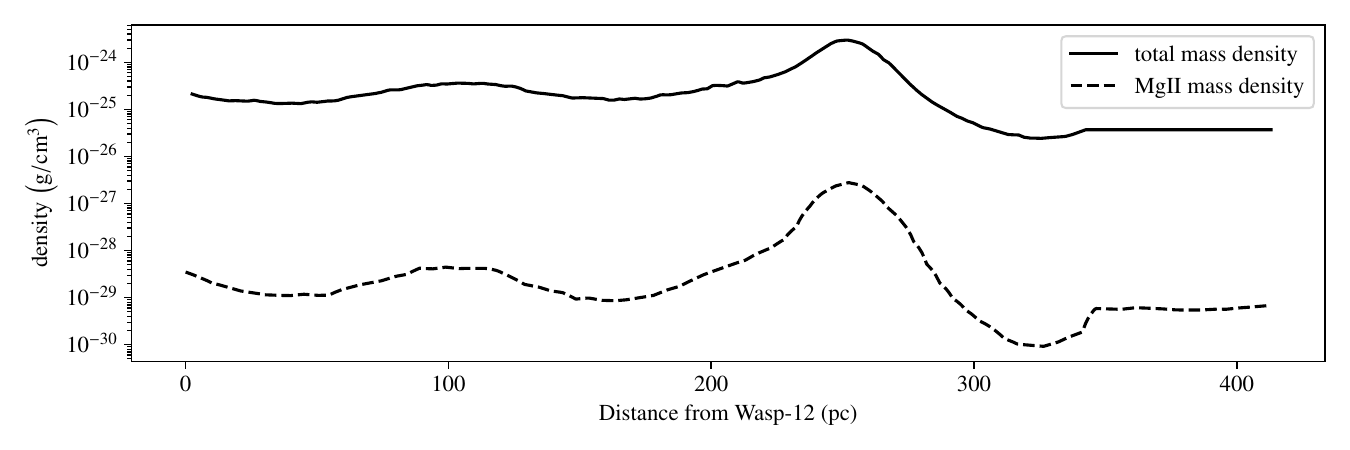}
	\caption[]{Total mass density and \ion{Mg}{ii} mass density profiles of the ISM along the LOS from WASP-12 (0 pc) to an observer on Earth (413 pc). These profile was derived using the methods described in Section \ref{sec:ism_profile_modelling}. The last $\sim$60 pc, in the vicinity of the Sun is not present in the extinction map of \cite{edenhofer2024ParsecscaleGalactic3D}, so we set the value to be constant at the closet value to the Sun, hence the flat profile here. The density in the region around the Sun is known to be lower than the average ISM density, so this flat profile maybe an overestimate, however it is still well below the average for the remaining parts of the profile and therefore should contribute negligibly to the overall column density.
  \label{fig:density_profile_ism}}
\end{figure*}

WASP-12 is located an estimated $413 \rm \ pc$ from Earth. The astrosphere simulation only extends to about 2 pc from WASP-12. In this Section, we analyse the density profile along the LOS of WASP-12 for the remaining $~ 411 \rm \ pc$, which has been derived based on the 3D dust extinction map around the Sun from \citet{edenhofer2024ParsecscaleGalactic3D} (see Section \ref{sec:ism_profile_modelling}). As can be seen in Fig. \ref{fig:density_profile_ism}, the hydrogen density profile of the ISM along the LOS is not uniform. Near WASP-12 the ISM has a density of around $2.22 \times 10^{-25} \rm \ g \ cm^{-3}$, which we have chosen as the initial density for the Astrosphere simulation in Section \ref{sec:astrosphere_sim}. The profile is dominated by a density peak roughly consistent with the edge of the Local Bubble \citep{cox1987LocalInterstellarMedium,linsky2021CouldLocalCavity,welsh2009TroubleLocalBubble,oneill2024LocalBubbleLocal} and falls off sharply at smaller distances. The map by \citet{edenhofer2024ParsecscaleGalactic3D} does not provide density measurements within 69 pc of the Sun, so we have assumed a constant density from the nearest value provided. As the density is known to be low in the solar neighbourhood, the exact choice of density will have a negligible impact on the total column density along the LOS. Integrating along the LOS gives a total mass column number density of $4.313\times 10^{-4} \ \mathrm{g} \cms$ (see Table \ref{tab:densities}) which is a factor of 3 lower than the $1.2\times 10^{-3} \ \mathrm{g} \cms$ assumed by \citet{Haswell2012}. 

\subsection{The full hydrogen profile and column density}
\label{sec:total_column_density}

\begin{figure*}
	\centering
	\includegraphics[width=0.99\textwidth,trim={0cm, 0cm, 0cm, 0cm},clip]{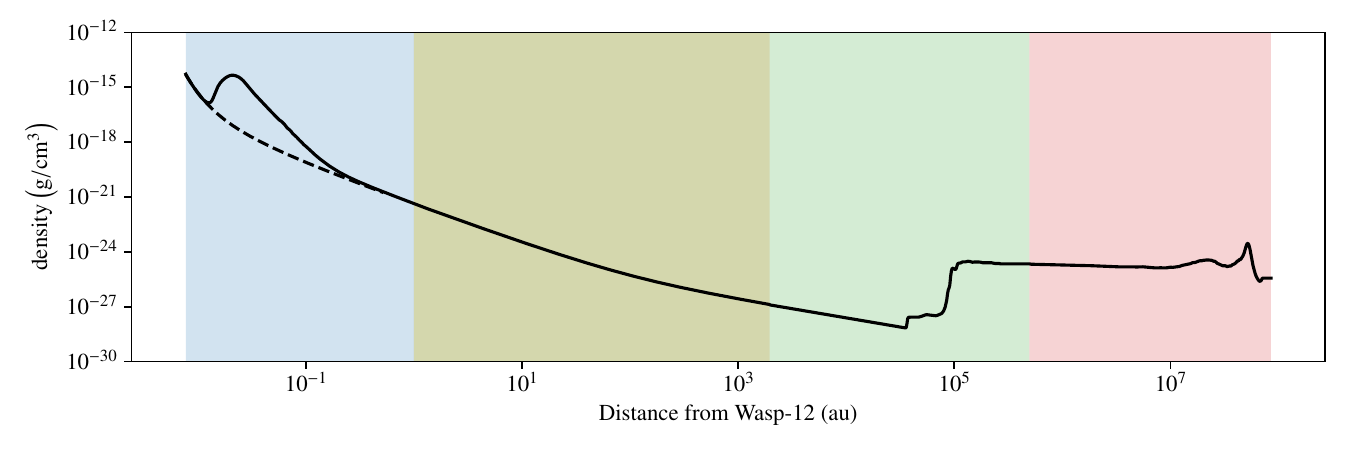}
	\caption[]{LOS hydrogen density profile from WASP-12 as seen from Earth. There are four distinct regions, highlighted in different colours from left to right, corresponding to the torus (blue), stellar wind (yellow), astrosphere (green) and ISM (red). There is also a dashed line indicating the wind profile in the absence of the torus, i.e. what the profile would be without the influence of the hot Jupiter. \label{fig:column_density_total}}
\end{figure*}

\begin{table*}
	\caption[]{Column densities for each component of the LOS density profile from WASP-12 to Earth.
	\label{tab:densities}}
	\centering
	\begin{tabular}{ccccc}
		\hline
		Column section & Mass density & \ion{H}{i} number density  & \ion{Mg}{ii} number density & \makecell{Percentage of total  {mass and} \\ {\ion{H}{i} number density}}  \\
		\hline
        torus       & 8.112$\times 10^{-4}$ g cm$^{-2}$ & 8.083$\times10^{20}$ cm$^{-2}$ & 6.534$\times10^{12}$ cm$^{-2}$ & 65.195\% \\
        ISM         & 4.313$\times 10^{-4}$ g cm$^{-2}$ & 4.298$\times10^{20}$ cm$^{-2}$ & 5.927$\times10^{15}$ cm$^{-2}$ & 34.662\% \\
        Astrosphere & 1.414$\times 10^{-6}$ g cm$^{-2}$ & 1.409$\times10^{18}$ cm$^{-2}$ & - & 0.114\% \\
        Parker wind & 2.405$\times 10^{-7}$ g cm$^{-2}$ & 2.396$\times10^{17}$ cm$^{-2}$ & - & 0.019\% \\
        \hline
        \hline
        Total & 1.244$\times 10^{-3}$ g cm$^{-2}$ & 1.240$\times10^{21}$ cm$^{-2}$ & 5.934$\times10^{15}$ cm$^{-2}$ & 100 \%\\
		\hline
	\end{tabular}
\end{table*}

In this section we combine the density profiles from the previous sections to piece together the full density profile from the surface of WASP-12 to an observer on Earth. To bridge the range of scales between the outer edge of the profile from the torus simulations in Section \ref{sec:torus_profile} to the inner edge of the astrosphere simulations in Section \ref{sec:astrosphere_profile} we extrapolate the profile from the torus simulations until it reaches the density profile expected from the stellar wind alone. This wind provides the inner boundary conditions for the astrosphere simulation, so no further interpolation is needed here. Similarly, the density profile from the astrosphere simulation has been interpolated onto the ISM near WASP-12, taken from the inner edge of the ISM profile in Section \ref{sec:ism_profile}. The full LOS density profile in hydrogen for WASP-12 is shown in Fig. \ref{fig:column_density_total}. 

As can be seen in Fig. \ref{fig:column_density_total}, the highest densities are found near the star. The gaseous torus fed by the planetary wind provides the densest feature along the profile, and even the inner stellar wind alone (dotted line) is denser than any other feature (See Section \ref{sec:torus_profile} for details). As it approaches the astrosphere, the stellar wind becomes under-dense in comparison to the local ISM. While difficult to see in this plot, some of this mass is swept up in the over-dense region at the edge of the astrosphere but due to the inclination angle of WASP-12, the densest gas in this region does not appear along the LOS (see Section \ref{sec:astrosphere_profile} for details). The full profile clearly shows that the dense cloud near the edge of the local bubble is significantly denser than any feature at the edge of the astrosphere but far more diffuse than the immediate environment of the star (see Section \ref{sec:ism_profile} for details).

Integrating along the LOS gives a total column density of $1.244\times10^3 \ \mathrm{g} \cms$, {which is equivalent to a  \ion{H}{I} number density of $1.24 \times 10^{21} \cms$ assuming solar metallicity and abundances along the entire line of sight.} Most of the column density comes from the region near the star, including the torus. The $10 \rm \ R_*$ near WASP-12, including the torus, contributes $~\sim 65\%$ to the total density. The second largest contributor is the ISM at $\sim 34\%$, due to its higher average densities and large spatial extent. The column density of both the extended Parker wind beyond $10 \rm \ R_*$ and the astrosphere are negligible, contributing only $0.1\%$ and $0.02\%$ respectively. The contributions of different components to the total hydrogen column density are summarised in Table \ref{tab:densities}.

If we assume an unperturbed Parker Wind from WASP-12 until the astrosphere, the total mass (\ion{H}{i} number) column density drops to $5.079 \times 10^{-4} \rm g \cms$ ($5.061 \times 10^{20} \cms$)  (dotted line in Fig. \ref{fig:column_density_total}). Therefore the LOS density enhancement due to the torus is $\sim 2.5$.

\subsection{Magnesium density along the line of sight}

To convert from gas density to Mg we assume a Mg abundance of $3.5\times 10^{-5}$ \citep{suess1956AbundancesElements}. Multiplying the total hydrogen column number density of $1.24\times 10^{21} \cms$ with the Mg abundance gives a total value for the Mg column density along the LOS from WASP-12 of $4.340\times10^{16} \cms$. Even assuming an extremely optimistic ionisation fraction of 1 this is lower ($\sim 4.6$ times lower) than the observationally required value of $2\times 10^{17} \cms$ reported by \citet{Haswell2012}. This means that even if Mg was fully ionised to \ion{Mg}{ii} all the way from WASP-12 to Earth, there is almost an order of magnitude too little gas to explain the observed complete absorption of the \ion{Mg}{ii} resonance lines.

{This would remain true even if we make optimistic assumptions: \citet{ehrenreich2011MasslossRatesTransiting} quote an upper end of the error bars of $2.7 \times 10^{12} \rm \ g \ s^{-1} $ for the planetary wind, a factor of 2.7 times higher than what we model here. If we assume that the resulting line-of-sight density in the torus also increased by the same amount then the total line of sight density would roughly double. This is still insufficient to explain the required line-of-sight density.}

\subsection{Estimating the ionization fraction of Mg and the MgII density along the line of sight}
\label{sec:torus_ionisation_fraction}

\begin{figure}
	\centering
	\includegraphics[width=0.49\textwidth,trim={0cm, 0cm, 0cm, 0cm},clip]{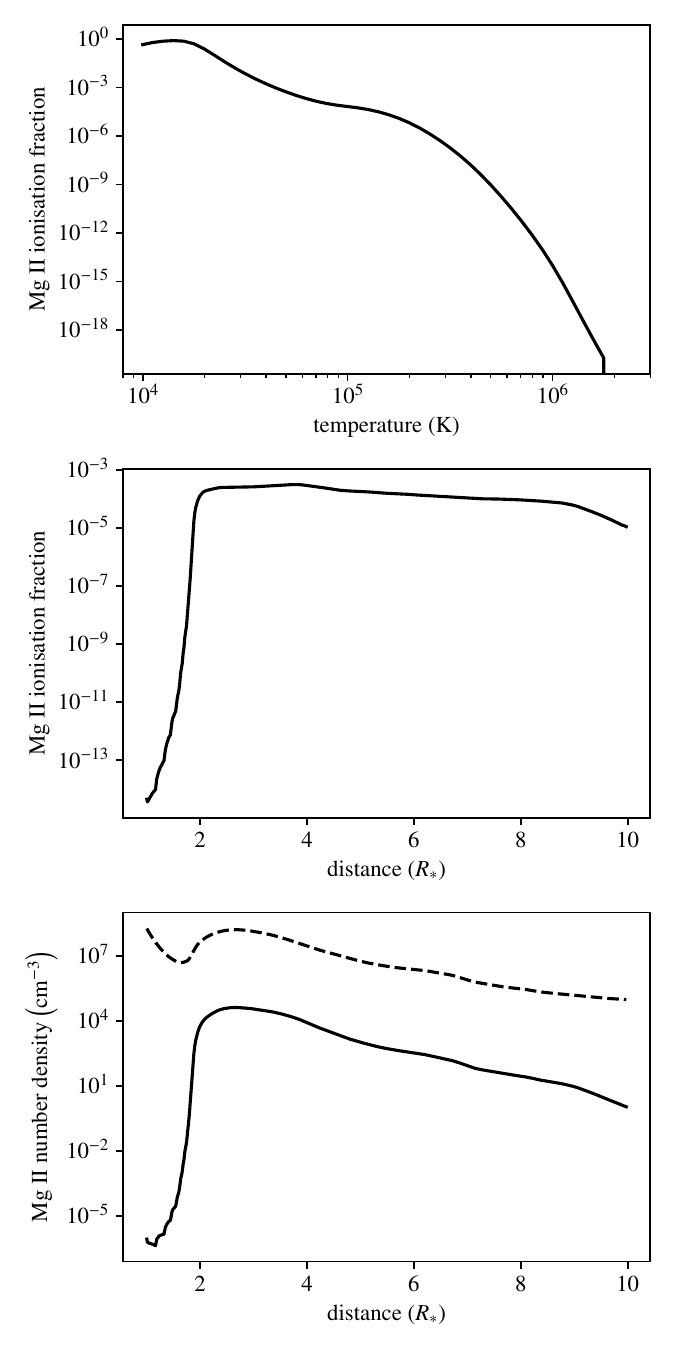}
	\caption[]{The top panel shows the ionisation fraction {of \ion{Mg}{ii}} as a function of temperature (top). Combining this with the  temperature profile from Fig. \ref{fig:torus_density}, we can show the \ion{Mg}{ii} ionisation fraction as a function of radius (middle). This allows us (Assuming a Mg abundance of $3.5\times10^{-4}$) to determine the resulting torus \ion{Mg}{ii} number density (bottom). We also include profile of \ion{Mg}{ii} that is calculated assuming that all of the Mg is ionised to \ion{Mg}{ii} (dashed line). All these radial profiles are along the LOS to the observer. \label{fig:ionization_fraction}}
\end{figure}

In this section we turn towards estimating the ionization fraction of \ion{Mg}{ii} for different components of the total column density, {i.e the fraction of \ion{Mg}{} that is singly ionised}.  As outlined in Section \ref{sec:total_column_density}, the contribution from gas in the extended Parker wind and at the edge of the astrosphere to the column density is negligible. For this reason we will focus only on the gas within the torus region and in the ISM. 

For the ISM, we can predict the LOS \ion{Mg}{ii} density using the 3D ionization map of \citet{mccallum2025HaSkyThree} (see Section \ref{sec:ism_profile_modelling} for details). This approach predicts a total \ion{Mg}{ii} column density of $5.927\times 10^{15} \cms$, which suggests an average ionization fraction of 0.394 along the LOS. This is somewhat lower than the 0.5 used in \citet{Haswell2012}. {As a combination of lower ionisation faction and lower gas density}, we predict that the ISM contribution is a factor of $\sim 3.3$ lower than the $2\times 10^{16} \cms$ assumed by \citet{Haswell2012}.
 
For the torus, we use the spectral synthesis code CHIANTIPy \citep{Dere1997, Dere2013, Dufresne2024}, along with the temperature profile of the torus. In the following we assume collisional ionization equilibrium. The top panel of Fig. \ref{fig:ionization_fraction} shows the ionization fraction due to thermal collisional ionization for \ion{Mg}{ii} as a function of temperature. While the ionization fractions are expected to be high for gas close to $10^4 \rm K$, they fall quickly with increasing temperature. This can be seen in the top panel of Fig. \ref{fig:ionization_fraction} (ionisation fraction as a function of temperature), as well as the bottom panel of Fig. \ref{fig:torus_density} (torus temperature). Even the coldest densest gas in the torus region along the LOS is found at a temperature closer to $5-6 \times 10^4 \ \rm K$. At these temperatures, the bulk of Mg is ionized to at least \ion{Mg}{iii} (and could be in a higher ionisation state). As a result of the temperature profile of the gas, the ionization fraction of gas from the torus onwards is of the order $10^{-4}$ to $10^{-5}$ and negligibly small for even smaller radii. The resulting number density of \ion{Mg}{ii}, shown in {the bottom panel} of Fig. \ref{fig:ionization_fraction}, is considerably lower than the optimistic estimate using an ionization fraction of 1 (dotted line). The resulting \ion{Mg}{ii} column density drops from $2.843\times10^{16} \cms$ to $6.534\times10^{12} \cms$ This means the overall contribution from the torus region to the LOS column density would decrease to only 0.015$\%$ of the total, while the total would drop to $5.934\times 10^{15} \cms$. This situation comes about because most of the Mg has been thermally ionised to \ion{Mg}{iii}. If the gas is optically thin, the additional photoionization from the background radiation field, that is responsible for ionizing the ISM, might push more \ion{Mg}{ii} into \ion{Mg}{iii}, further decreasing the \ion{Mg}{ii} column density.

{An important consideration is that the LOS velocity of the WASP-12 system is different from the LOS velocity of the ISM. This difference means that any ISM absorption features should be blue-shifted with respect the WASP-12 features.
\cite{Czesla2024} (section 3.4, Table 5) confirms earlier measurements by \cite{Fossati2013} of the velocity of two ISM components in the \ion{Na}{I} D$_{2}$ lines. There are apparently 2 components, both blue-shifted relative to the spectrum of WASP-12. Their barycentric RVs are 2.75 and 12.7 km/s respectively; in the same frame, WASP-12 recedes at 19.46 km/s. Hence, the ISM absorption components are blue-shifted by -16.7 and -6.76 km/s relative to WASP-12. This makes an interstellar origin for the \ion{Mg}{II} absorption less plausible, a point made previously by \cite{Fossati2013}.}

A caveat to our conclusions is that we have assumed that cooling within the torus is negligible. Our results show that most of the gas in the torus is currently found near $10^5$ K. At this temperature, optically thin radiative cooling is highly efficient and the gas can radiate energy away rapidly and cool to temperatures of the order of the effective temperature of the star, $10^4 K$ or lower, where ionization fractions rapidly rise towards unity (see top right panel of Fig. \ref{fig:ionization_fraction}). Cooler gas results in higher ionization fractions for \ion{Mg}{ii} as gas is no longer triply ionized to \ion{Mg}{iii}. Under these conditions, photo-ionization from nearby O- and B-stars could also become important if the gas is optically thin; however, at these temperatures, the gas is closer to conditions seen in solar and stellar prominences, where optically thick effects become important. 

Cooling would lead to fragmentation via the cooling instability (see \cite{Hermans2021} for details on this phenomenon), resulting in smaller clouds of higher densities. This would result in a torus with a multi-phase gas structure. This mix of gas states could delay the convergence of the torus mass and result in higher densities than are seen in Fig. \ref{fig:mass_time}, allowing more mass to build up. One intriguing possibility is that a cooling, fragmenting torus could rain down onto the surface of WASP-12 as dense cool clumps of planetary material. This would form a type of planetary coronal rain, a paradigm which was investigated by \cite{DaleyYates2019} for the system HD 189733. 

We will leave investigating torus cooling and fragmentation to future work, which directly models the balance between optically thin radiative cooling and the heating of the stellar wind via coronal physics.

If cooling leads to fragmentation, one could also possibly expect a changing covering fraction, which might be detectable in absorption values that vary over time. This may also explain the lack of torus detection by \cite{Czesla2024}. 

\section{Conclusions}
\label{sec:conclusions}
Observations of the star WASP-12 shows strong absorption in the \ion{Mg}{ii} resonance absorption line. In this paper we investigated whether the total column density of \ion{Mg}{ii} along the LOS is sufficient to explain the observed absorption. 

To estimate the full density profile of \ion{Mg}{ii} along the LOS, we combined a set of hydrodynamical simulations with observations of the interstellar medium. Our work shows that there is insufficient gas density along the LOS to completely explain the reported absorption in the \ion{Mg}{ii} line. Even when assuming Mg is fully ionised into \ion{Mg}{ii}, we only report a total column density of $3.805\times 10^{16} \cms$, which falls significantly short of the required $2\times 10^{17} \cms$ required by \cite{Haswell2012}. With physically motivated assumptions for the ionisation, the \ion{Mg}{ii} column density drops to only $5.934\times 10^{15} \cms$. Specifically, we report that:
\begin{itemize}
\item The bulk of the hydrogen column density is located within the dense torus that is fed by the wind from the planet WASP-12b. We show for the first time that the density profile of the torus saturates after only 60 orbits, following the linear growth phase reported in \cite{Debrecht2018}. Peak densities within the torus reach $\sim 5\times10^{-15} \rm \ g \ cm^{-3}$ and the region covered by the torus contributes up to $8.083 \times 10^{20} \cms$ to the column number density along the LOS. 
\item Most of the gas in the torus is sufficiently hot to be ionised to \ion{Mg}{iii}. As a result we estimate that the total column number density of \ion{Mg}{ii} supplied by the torus region is only $6.534 \times 10^{12} \cms$, 0.015 percent of the \ion{Mg}{ii} column density along the LOS. 
\item The gas from the outer edge of the torus to the astrosphere of WASP-12 along the LOS contributes a negligible fraction (0.133 percent) of the total column density and negligibly to the \ion{Mg}{ii} column density.
\item The ISM contributes $\sim34\%$ of the gas column density. We estimate that the average ionization fraction of \ion{Mg}{ii} in the ISM along the LOS to WASP-12 is 0.394, due to photoionization by a collection of nearby O- and B-stars. As a result, the ISM contributes $99.985\%$ of the \ion{Mg}{ii} column density along the LOS.
\end{itemize}

One caveat to the work presented here is that the final state of our torus simulation predicts that the gas should be strongly cooling, a physical process that has been neglected here. If this cooling could delay the saturation of the torus it might be able to significantly increase the column density of gas contained within the torus, albeit likely at the cost of a clumpy torus structure with a time-varying covering fraction. A cooler torus would also potentially significantly increase the ionization fraction of \ion{Mg}{ii} in the torus, which could have a significant impact on the \ion{Mg}{ii} column density along the LOS due to the large amount of gas mass contained in this region. We leave this investigation to future work.

\section*{Data availability}
A CSV file of the data presented in Fig. \ref{fig:column_density_total} is provided online.

\section{Acknowledgements}

{The authors thank the reviewer for their comments and suggestions.}

SD-Y and MJ acknowledge support from STFC consolidated grant number ST/R000824/1. RSB acknowledges support from a UKRI Future Leaders Fellowship (grant code: MR/Y015517/1). For the purpose of open access, the authors have applied a Creative Commons Attribution (CC BY) licence to any Author Accepted Manuscript version arising.
	
\section*{ORICD iDs}

Simon Daley-Yates \orcidlink{0000-0002-0461-3029} \url{https://orcid.org/0000-0002-0461-3029} \\
Ricarda S. Beckmann \orcidlink{0000-0002-2850-0192} \url{https://orcid.org/0000-0002-2850-0192} \\
Lewis McCallum \orcidlink{0009-0007-1181-8034} \url{https://orcid.org/0009-0007-1181-8034} \\
Moira M. Jardine \orcidlink{0000-0002-1466-5236} \url{https://orcid.org/0000-0002-1466-5236} \\
Andrew C. Cameron \orcidlink{0000-0002-8863-7828} \url{https://orcid.org/0000-0002-8863-7828}

\bibliographystyle{mnras}
\bibliography{references,ricarda_WASP12_references}

@article{suess1956AbundancesElements,
  title = {Abundances of the {{Elements}}},
  author = {Suess, Hans E. and Urey, Harold C.},
  year = {1956},
  month = jan,
  journal = {Reviews of Modern Physics},
  volume = {28},
  pages = {53--74},
  publisher = {APS},
  issn = {0034-6861},
  doi = {10.1103/RevModPhys.28.53},
  urldate = {2025-07-10}
}

@article{ehrenreich2011MasslossRatesTransiting,
  title = {Mass-Loss Rates for Transiting Exoplanets},
  author = {Ehrenreich, D. and D{\'e}sert, J. -M.},
  year = {2011},
  month = may,
  volume = {529},
  pages = {A136},
  issn = {0004-6361},
  doi = {10.1051/0004-6361/201016356},
  urldate = {2025-07-09},
  journal = {\aap}
}

@article{guo2011EscapingParticleFluxes,
  title = {Escaping {{Particle Fluxes}} in the {{Atmospheres}} of {{Close-in Exoplanets}}. {{I}}. {{Model}} of {{Hydrogen}}},
  author = {Guo, J. H.},
  year = {2011},
  month = jun,
  volume = {733},
  pages = {98},
  publisher = {IOP},
  issn = {0004-637X},
  doi = {10.1088/0004-637X/733/2/98},
  urldate = {2025-07-09},
  journal = {\apj}
}

@article{hebb2009WASP12bHottestTransiting,
  title = {{{WASP-12b}}: {{The Hottest Transiting Extrasolar Planet Yet Discovered}}},
  shorttitle = {{{WASP-12b}}},
  author = {Hebb, L. and {Collier-Cameron}, A. and Loeillet, B. and Pollacco, D. and H{\'e}brard, G. and Street, R. A. and Bouchy, F. and Stempels, H. C. and Moutou, C. and Simpson, E. and Udry, S. and Joshi, Y. C. and West, R. G. and Skillen, I. and Wilson, D. M. and McDonald, I. and Gibson, N. P. and Aigrain, S. and Anderson, D. R. and Benn, C. R. and Christian, D. J. and Enoch, B. and Haswell, C. A. and Hellier, C. and Horne, K. and Irwin, J. and Lister, T. A. and Maxted, P. and Mayor, M. and Norton, A. J. and Parley, N. and Pont, F. and Queloz, D. and Smalley, B. and Wheatley, P. J.},
  year = {2009},
  month = mar,
  volume = {693},
  pages = {1920--1928},
  publisher = {IOP},
  issn = {0004-637X},
  doi = {10.1088/0004-637X/693/2/1920},
  urldate = {2025-07-09},
  journal = {\apj}
}

@article{husnoo2011OrbitalEccentricityWASP12,
  title = {Orbital Eccentricity of {{WASP-12}} and {{WASP-14}} from New Radial Velocity Monitoring with {{SOPHIE}}},
  author = {Husnoo, Nawal and Pont, Fr{\'e}d{\'e}ric and H{\'e}brard, Guillaume and Simpson, Elaine and Mazeh, Tsevi and Bouchy, Fran{\c c}ois and Moutou, Claire and Arnold, Luc and Boisse, Isabelle and D{\'i}az, Rodrigo F. and Eggenberger, Anne and Shporer, Avi},
  year = {2011},
  month = jun,
  volume = {413},
  pages = {2500--2508},
  publisher = {OUP},
  issn = {0035-8711},
  doi = {10.1111/j.1365-2966.2011.18322.x},
  urldate = {2025-07-09},
  journal = {\mnras}
}

@article{li2010WASP12bProlateInflated,
  title = {{{WASP-12b}} as a Prolate, Inflated and Disrupting Planet from Tidal Dissipation},
  author = {Li, Shu-Lin and Miller, N. and Lin, Douglas N. C. and Fortney, Jonathan J.},
  year = {2010},
  month = feb,
  journal = {Nature},
  volume = {463},
  pages = {1054--1056},
  issn = {0028-0836},
  doi = {10.1038/nature08715},
  urldate = {2025-07-09}
}

@article{llama2011ShockingTransitWASP12b,
  title = {The Shocking Transit of {{WASP-12b}}: Modelling the Observed Early Ingress in the near-Ultraviolet},
  shorttitle = {The Shocking Transit of {{WASP-12b}}},
  author = {Llama, J. and Wood, K. and Jardine, M. and Vidotto, A. A. and Helling, {\relax Ch}. and Fossati, L. and Haswell, C. A.},
  year = {2011},
  month = sep,
  volume = {416},
  pages = {L41-L44},
  publisher = {OUP},
  issn = {0035-8711},
  doi = {10.1111/j.1745-3933.2011.01093.x},
  urldate = {2025-07-09},
  journal = {\mnras}
}

@article{lopez-morales2010DaysideBandEmission,
  title = {Day-Side z'-Band {{Emission}} and {{Eccentricity}} of {{WASP-12b}}},
  author = {{L{\'o}pez-Morales}, Mercedes and Coughlin, Jeffrey L. and Sing, David K. and Burrows, Adam and Apai, D{\'a}niel and Rogers, Justin C. and Spiegel, David S. and Adams, Elisabeth R.},
  year = {2010},
  month = jun,
  volume = {716},
  pages = {L36-L40},
  publisher = {IOP},
  issn = {0004-637X},
  doi = {10.1088/2041-8205/716/1/L36},
  urldate = {2025-07-09},
  journal = {\apj}
}

@article{vidal-madjar2004DetectionOxygenCarbon,
  title = {Detection of {{Oxygen}} and {{Carbon}} in the {{Hydrodynamically Escaping Atmosphere}} of the {{Extrasolar Planet HD}} 209458b},
  author = {{Vidal-Madjar}, A. and D{\'e}sert, J. -M. and {Lecavelier des Etangs}, A. and H{\'e}brard, G. and Ballester, G. E. and Ehrenreich, D. and Ferlet, R. and McConnell, J. C. and Mayor, M. and Parkinson, C. D.},
  year = {2004},
  month = mar,
  volume = {604},
  pages = {L69-L72},
  publisher = {IOP},
  issn = {0004-637X},
  doi = {10.1086/383347},
  urldate = {2025-07-09},
  journal = {\apj}
}

@article{vidotto2010EarlyUVIngress,
  title = {Early {{UV Ingress}} in {{WASP-12b}}: {{Measuring Planetary Magnetic Fields}}},
  shorttitle = {Early {{UV Ingress}} in {{WASP-12b}}},
  author = {Vidotto, A. A. and Jardine, M. and Helling, {\relax Ch}.},
  year = {2010},
  month = oct,
  volume = {722},
  pages = {L168-L172},
  publisher = {IOP},
  issn = {0004-637X},
  doi = {10.1088/2041-8205/722/2/L168},
  urldate = {2025-07-09},
  journal = {\apj}
}

@ARTICLE{Bonomo2017,
       author = {{Bonomo}, A.~S. and {Desidera}, S. and {Benatti}, S. and {Borsa}, F. and {Crespi}, S. and {Damasso}, M. and {Lanza}, A.~F. and {Sozzetti}, A. and {Lodato}, G. and {Marzari}, F. and {Boccato}, C. and {Claudi}, R.~U. and {Cosentino}, R. and {Covino}, E. and {Gratton}, R. and {Maggio}, A. and {Micela}, G. and {Molinari}, E. and {Pagano}, I. and {Piotto}, G. and {Poretti}, E. and {Smareglia}, R. and {Affer}, L. and {Biazzo}, K. and {Bignamini}, A. and {Esposito}, M. and {Giacobbe}, P. and {H{\'e}brard}, G. and {Malavolta}, L. and {Maldonado}, J. and {Mancini}, L. and {Martinez Fiorenzano}, A. and {Masiero}, S. and {Nascimbeni}, V. and {Pedani}, M. and {Rainer}, M. and {Scandariato}, G.},
        title = "{The GAPS Programme with HARPS-N at TNG . XIV. Investigating giant planet migration history via improved eccentricity and mass determination for 231 transiting planets}",
      journal = {\aap},
         year = 2017,
        month = jun,
       volume = {602},
          eid = {A107},
        pages = {A107},
          doi = {10.1051/0004-6361/201629882},
archivePrefix = {arXiv},
       eprint = {1704.00373},
 primaryClass = {astro-ph.EP},
       adsurl = {https://ui.adsabs.harvard.edu/abs/2017A&A...602A.107B}
}

@ARTICLE{Fossati2010,
       author = {{Fossati}, L. and {Bagnulo}, S. and {Elmasli}, A. and {Haswell}, C.~A. and {Holmes}, S. and {Kochukhov}, O. and {Shkolnik}, E.~L. and {Shulyak}, D.~V. and {Bohlender}, D. and {Albayrak}, B. and {Froning}, C. and {Hebb}, L.},
        title = "{A Detailed Spectropolarimetric Analysis of the Planet-hosting Star WASP-12}",
      journal = {\apj},
         year = 2010,
        month = sep,
       volume = {720},
       number = {1},
        pages = {872-886},
          doi = {10.1088/0004-637X/720/1/872},
archivePrefix = {arXiv},
       eprint = {1007.3082},
 primaryClass = {astro-ph.SR},
       adsurl = {https://ui.adsabs.harvard.edu/abs/2010ApJ...720..872F}
}

@ARTICLE{Fossati2013,
       author = {{Fossati}, L. and {Ayres}, T.~R. and {Haswell}, C.~A. and {Bohlender}, D. and {Kochukhov}, O. and {Fl{\"o}er}, L.},
        title = "{Absorbing Gas around the WASP-12 Planetary System}",
      journal = {\apjl},
         year = 2013,
        month = apr,
       volume = {766},
       number = {2},
          eid = {L20},
        pages = {L20},
          doi = {10.1088/2041-8205/766/2/L20},
archivePrefix = {arXiv},
       eprint = {1303.3375},
 primaryClass = {astro-ph.SR},
       adsurl = {https://ui.adsabs.harvard.edu/abs/2013ApJ...766L..20F}
}

@ARTICLE{Haswell2012,
       author = {{Haswell}, C.~A. and {Fossati}, L. and {Ayres}, T. and {France}, K. and {Froning}, C.~S. and {Holmes}, S. and {Kolb}, U.~C. and {Busuttil}, R. and {Street}, R.~A. and {Hebb}, L. and {Collier Cameron}, A. and {Enoch}, B. and {Burwitz}, V. and {Rodriguez}, J. and {West}, R.~G. and {Pollacco}, D. and {Wheatley}, P.~J. and {Carter}, A.},
        title = "{Near-ultraviolet Absorption, Chromospheric Activity, and Star-Planet Interactions in the WASP-12 system}",
      journal = {\apj},
         year = 2012,
        month = nov,
       volume = {760},
       number = {1},
          eid = {79},
        pages = {79},
          doi = {10.1088/0004-637X/760/1/79},
archivePrefix = {arXiv},
       eprint = {1301.1860},
 primaryClass = {astro-ph.EP},
       adsurl = {https://ui.adsabs.harvard.edu/abs/2012ApJ...760...79H}
}

@ARTICLE{Debrecht2018,
       author = {{Debrecht}, Alex and {Carroll-Nellenback}, Jonathan and {Frank}, Adam and {Fossati}, Luca and {Blackman}, Eric G. and {Dobbs-Dixon}, Ian},
        title = "{Generation of a circumstellar gas disc by hot Jupiter WASP-12b}",
      journal = {\mnras},
         year = 2018,
        month = aug,
       volume = {478},
       number = {2},
        pages = {2592-2598},
          doi = {10.1093/mnras/sty1164},
archivePrefix = {arXiv},
       eprint = {1805.00596},
 primaryClass = {astro-ph.EP},
       adsurl = {https://ui.adsabs.harvard.edu/abs/2018MNRAS.478.2592D}
}

@ARTICLE{DaleyYates2019,
       author = {{Daley-Yates}, S. and {Stevens}, I.~R.},
        title = "{Hot Jupiter accretion: 3D MHD simulations of star-planet-wind interaction}",
      journal = {\mnras},
         year = 2019,
        month = feb,
       volume = {483},
       number = {2},
        pages = {2600-2614},
          doi = {10.1093/mnras/sty3310},
archivePrefix = {arXiv},
       eprint = {1812.00665},
 primaryClass = {astro-ph.EP},
       adsurl = {https://ui.adsabs.harvard.edu/abs/2019MNRAS.483.2600D}
}

@ARTICLE{Mignone2007,
       author = {{Mignone}, A. and {Bodo}, G. and {Massaglia}, S. and {Matsakos}, T. and {Tesileanu}, O. and {Zanni}, C. and {Ferrari}, A.},
        title = "{PLUTO: A Numerical Code for Computational Astrophysics}",
      journal = {\apjs},
         year = 2007,
        month = may,
       volume = {170},
       number = {1},
        pages = {228-242},
          doi = {10.1086/513316},
archivePrefix = {arXiv},
       eprint = {astro-ph/0701854},
 primaryClass = {astro-ph},
       adsurl = {https://ui.adsabs.harvard.edu/abs/2007ApJS..170..228M}
}

@ARTICLE{Binzheng2019,
       author = {{Zhang}, Binzheng and {Sorathia}, Kareem A. and {Lyon}, John G. and {Merkin}, Viacheslav G. and {Wiltberger}, Michael},
        title = "{Conservative averaging-reconstruction techniques (Ring Average) for 3-D finite-volume MHD solvers with axis singularity}",
      journal = {Journal of Computational Physics},
         year = 2019,
        month = jan,
       volume = {376},
        pages = {276-294},
          doi = {10.1016/j.jcp.2018.08.020},
       adsurl = {https://ui.adsabs.harvard.edu/abs/2019JCoPh.376..276Z}
}

@ARTICLE{Wood2005,
       author = {{Wood}, B.~E. and {M{\"u}ller}, H. -R. and {Zank}, G.~P. and {Linsky}, J.~L. and {Redfield}, S.},
        title = "{New Mass-Loss Measurements from Astrospheric Ly{\ensuremath{\alpha}} Absorption}",
      journal = {\apjl},
         year = 2005,
        month = aug,
       volume = {628},
       number = {2},
        pages = {L143-L146},
          doi = {10.1086/432716},
archivePrefix = {arXiv},
       eprint = {astro-ph/0506401},
 primaryClass = {astro-ph},
       adsurl = {https://ui.adsabs.harvard.edu/abs/2005ApJ...628L.143W}
}

@ARTICLE{Gaia32023,
       author = {{Gaia Collaboration} and {Vallenari}, A. and {Brown}, A.~G.~A. and {Prusti}, T. and {de Bruijne}, J.~H.~J. and {Arenou}, F. and {Babusiaux}, C. and {Biermann}, M. and {Creevey}, O.~L. and {Ducourant}, C. and {Evans}, D.~W. and {Eyer}, L. and {Guerra}, R. and {Hutton}, A. and {Jordi}, C. and {Klioner}, S.~A. and {Lammers}, U.~L. and {Lindegren}, L. and {Luri}, X. and {Mignard}, F. and {Panem}, C. and {Pourbaix}, D. and {Randich}, S. and {Sartoretti}, P. and {Soubiran}, C. and {Tanga}, P. and {Walton}, N.~A. and {Bailer-Jones}, C.~A.~L. and {Bastian}, U. and {Drimmel}, R. and {Jansen}, F. and {Katz}, D. and {Lattanzi}, M.~G. and {van Leeuwen}, F. and {Bakker}, J. and {Cacciari}, C. and {Casta{\~n}eda}, J. and {De Angeli}, F. and {Fabricius}, C. and {Fouesneau}, M. and {Fr{\'e}mat}, Y. and {Galluccio}, L. and {Guerrier}, A. and {Heiter}, U. and {Masana}, E. and {Messineo}, R. and {Mowlavi}, N. and {Nicolas}, C. and {Nienartowicz}, K. and {Pailler}, F. and {Panuzzo}, P. and {Riclet}, F. and {Roux}, W. and {Seabroke}, G.~M. and {Sordo}, R. and {Th{\'e}venin}, F. and {Gracia-Abril}, G. and {Portell}, J. and {Teyssier}, D. and {Altmann}, M. and {Andrae}, R. and {Audard}, M. and {Bellas-Velidis}, I. and {Benson}, K. and {Berthier}, J. and {Blomme}, R. and {Burgess}, P.~W. and {Busonero}, D. and {Busso}, G. and {C{\'a}novas}, H. and {Carry}, B. and {Cellino}, A. and {Cheek}, N. and {Clementini}, G. and {Damerdji}, Y. and {Davidson}, M. and {de Teodoro}, P. and {Nu{\~n}ez Campos}, M. and {Delchambre}, L. and {Dell'Oro}, A. and {Esquej}, P. and {Fern{\'a}ndez-Hern{\'a}ndez}, J. and {Fraile}, E. and {Garabato}, D. and {Garc{\'\i}a-Lario}, P. and {Gosset}, E. and {Haigron}, R. and {Halbwachs}, J. -L. and {Hambly}, N.~C. and {Harrison}, D.~L. and {Hern{\'a}ndez}, J. and {Hestroffer}, D. and {Hodgkin}, S.~T. and {Holl}, B. and {Jan{\ss}en}, K. and {Jevardat de Fombelle}, G. and {Jordan}, S. and {Krone-Martins}, A. and {Lanzafame}, A.~C. and {L{\"o}ffler}, W. and {Marchal}, O. and {Marrese}, P.~M. and {Moitinho}, A. and {Muinonen}, K. and {Osborne}, P. and {Pancino}, E. and {Pauwels}, T. and {Recio-Blanco}, A. and {Reyl{\'e}}, C. and {Riello}, M. and {Rimoldini}, L. and {Roegiers}, T. and {Rybizki}, J. and {Sarro}, L.~M. and {Siopis}, C. and {Smith}, M. and {Sozzetti}, A. and {Utrilla}, E. and {van Leeuwen}, M. and {Abbas}, U. and {{\'A}brah{\'a}m}, P. and {Abreu Aramburu}, A. and {Aerts}, C. and {Aguado}, J.~J. and {Ajaj}, M. and {Aldea-Montero}, F. and {Altavilla}, G. and {{\'A}lvarez}, M.~A. and {Alves}, J. and {Anders}, F. and {Anderson}, R.~I. and {Anglada Varela}, E. and {Antoja}, T. and {Baines}, D. and {Baker}, S.~G. and {Balaguer-N{\'u}{\~n}ez}, L. and {Balbinot}, E. and {Balog}, Z. and {Barache}, C. and {Barbato}, D. and {Barros}, M. and {Barstow}, M.~A. and {Bartolom{\'e}}, S. and {Bassilana}, J. -L. and {Bauchet}, N. and {Becciani}, U. and {Bellazzini}, M. and {Berihuete}, A. and {Bernet}, M. and {Bertone}, S. and {Bianchi}, L. and {Binnenfeld}, A. and {Blanco-Cuaresma}, S. and {Blazere}, A. and {Boch}, T. and {Bombrun}, A. and {Bossini}, D. and {Bouquillon}, S. and {Bragaglia}, A. and {Bramante}, L. and {Breedt}, E. and {Bressan}, A. and {Brouillet}, N. and {Brugaletta}, E. and {Bucciarelli}, B. and {Burlacu}, A. and {Butkevich}, A.~G. and {Buzzi}, R. and {Caffau}, E. and {Cancelliere}, R. and {Cantat-Gaudin}, T. and {Carballo}, R. and {Carlucci}, T. and {Carnerero}, M.~I. and {Carrasco}, J.~M. and {Casamiquela}, L. and {Castellani}, M. and {Castro-Ginard}, A. and {Chaoul}, L. and {Charlot}, P. and {Chemin}, L. and {Chiaramida}, V. and {Chiavassa}, A. and {Chornay}, N. and {Comoretto}, G. and {Contursi}, G. and {Cooper}, W.~J. and {Cornez}, T. and {Cowell}, S. and {Crifo}, F. and {Cropper}, M. and {Crosta}, M. and {Crowley}, C. and {Dafonte}, C. and {Dapergolas}, A. and {David}, M. and {David}, P. and {de Laverny}, P. and {De Luise}, F. and {De March}, R.},
        title = "{Gaia Data Release 3. Summary of the content and survey properties}",
      journal = {\aap},
         year = 2023,
        month = jun,
       volume = {674},
          eid = {A1},
        pages = {A1},
          doi = {10.1051/0004-6361/202243940},
archivePrefix = {arXiv},
       eprint = {2208.00211},
 primaryClass = {astro-ph.GA},
       adsurl = {https://ui.adsabs.harvard.edu/abs/2023A&A...674A...1G}
}

@ARTICLE{Czesla2024,
       author = {{Czesla}, S. and {Lamp{\'o}n}, M. and {Cont}, D. and {Lesjak}, F. and {Orell-Miquel}, J. and {Sanz-Forcada}, J. and {Nagel}, E. and {Nortmann}, L. and {Molaverdikhani}, K. and {L{\'o}pez-Puertas}, M. and {Yan}, F. and {Quirrenbach}, A. and {Caballero}, J.~A. and {Pall{\'e}}, E. and {Aceituno}, J. and {Amado}, P.~J. and {Henning}, Th. and {Khalafinejad}, S. and {Montes}, D. and {Reiners}, A. and {Ribas}, I. and {Schweitzer}, A.},
        title = "{The elusive atmosphere of WASP-12 b. High-resolution transmission spectroscopy with CARMENES}",
      journal = {\aap},
         year = 2024,
        month = mar,
       volume = {683},
          eid = {A67},
        pages = {A67},
          doi = {10.1051/0004-6361/202348107},
archivePrefix = {arXiv},
       eprint = {2401.02195},
 primaryClass = {astro-ph.EP},
       adsurl = {https://ui.adsabs.harvard.edu/abs/2024A&A...683A..67C}
}

@ARTICLE{Lai2010,
       author = {{Lai}, Dong and {Helling}, Ch. and {van den Heuvel}, E.~P.~J.},
        title = "{Mass Transfer, Transiting Stream, and Magnetopause in Close-in Exoplanetary Systems with Applications to WASP-12}",
      journal = {\apj},
         year = 2010,
        month = oct,
       volume = {721},
       number = {2},
        pages = {923-928},
          doi = {10.1088/0004-637X/721/2/923},
archivePrefix = {arXiv},
       eprint = {1005.4497},
 primaryClass = {astro-ph.EP},
       adsurl = {https://ui.adsabs.harvard.edu/abs/2010ApJ...721..923L}
}

@ARTICLE{Chakrabarty2019,
       author = {{Chakrabarty}, Aritra and {Sengupta}, Sujan},
        title = "{Precise Photometric Transit Follow-up Observations of Five Close-in Exoplanets: Update on Their Physical Properties}",
      journal = {\aj},
         year = 2019,
        month = jul,
       volume = {158},
       number = {1},
          eid = {39},
        pages = {39},
          doi = {10.3847/1538-3881/ab24dd},
archivePrefix = {arXiv},
       eprint = {1905.11258},
 primaryClass = {astro-ph.EP},
       adsurl = {https://ui.adsabs.harvard.edu/abs/2019AJ....158...39C}
}

@ARTICLE{Ozturk2019,
       author = {{{\"O}zt{\"u}rk}, O{\v{g}}uz and {Erdem}, Ahmet},
        title = "{New photometric analysis of five exoplanets: CoRoT-2b, HAT-P-12b, TrES-2b, WASP-12b, and WASP-52b}",
      journal = {\mnras},
         year = 2019,
        month = jun,
       volume = {486},
       number = {2},
        pages = {2290-2307},
          doi = {10.1093/mnras/stz747},
       adsurl = {https://ui.adsabs.harvard.edu/abs/2019MNRAS.486.2290O}
}

@article{Hermans2021,
	adsurl = {https://ui.adsabs.harvard.edu/abs/2021A\&A...655A..36H},
	archiveprefix = {arXiv},
	author = {{Hermans}, J. and {Keppens}, R.},
	doi = {10.1051/0004-6361/202140665},
	eid = {A36},
	eprint = {2107.07569},
	journal = {\aap},
	month = nov,
	pages = {A36},
	primaryclass = {astro-ph.SR},
	title = {Effect of optically thin cooling curves on condensation formation: Case study using thermal instability},
	volume = {655},
	year = 2021
}

@ARTICLE{Parker1965,
       author = {{Parker}, E.~N.},
        title = "{Dynamical Theory of the Solar Wind}",
      journal = {\ssr},
         year = 1965,
        month = sep,
       volume = {4},
       number = {5-6},
        pages = {666-708},
          doi = {10.1007/BF00216273},
       adsurl = {https://ui.adsabs.harvard.edu/abs/1965SSRv....4..666P}
}

@ARTICLE{Osterman2011,
       author = {{Osterman}, S. and {Green}, J. and {Froning}, C. and {B{\'e}land}, S. and {Burgh}, E. and {France}, K. and {Penton}, S. and {Delker}, T. and {Ebbets}, D. and {Sahnow}, D. and {Bacinski}, J. and {Kimble}, R. and {Andrews}, J. and {Wilkinson}, E. and {McPhate}, J. and {Siegmund}, O. and {Ake}, T. and {Aloisi}, A. and {Biagetti}, C. and {Diaz}, R. and {Dixon}, W. and {Friedman}, S. and {Ghavamian}, P. and {Goudfrooij}, P. and {Hartig}, G. and {Keyes}, C. and {Lennon}, D. and {Massa}, D. and {Niemi}, S. and {Oliveira}, C. and {Osten}, R. and {Proffitt}, C. and {Smith}, T. and {Soderblom}, D.},
        title = "{The Cosmic Origins Spectrograph: on-orbit instrument performance}",
      journal = {\apss},
         year = 2011,
        month = sep,
       volume = {335},
       number = {1},
        pages = {257-265},
          doi = {10.1007/s10509-011-0699-5},
archivePrefix = {arXiv},
       eprint = {1012.5827},
 primaryClass = {astro-ph.IM},
       adsurl = {https://ui.adsabs.harvard.edu/abs/2011Ap&SS.335..257O}
}

@ARTICLE{Leonardi2024,
       author = {{Leonardi}, P. and {Nascimbeni}, V. and {Granata}, V. and {Malavolta}, L. and {Borsato}, L. and {Biazzo}, K. and {Lanza}, A.~F. and {Desidera}, S. and {Piotto}, G. and {Nardiello}, D. and {Damasso}, M. and {Cunial}, A. and {Bedin}, L.~R.},
        title = "{TASTE. V. A new ground-based investigation of orbital decay in the ultra-hot Jupiter WASP-12b}",
      journal = {\aap},
         year = 2024,
        month = jun,
       volume = {686},
          eid = {A84},
        pages = {A84},
          doi = {10.1051/0004-6361/202348363},
archivePrefix = {arXiv},
       eprint = {2402.12120},
 primaryClass = {astro-ph.EP},
       adsurl = {https://ui.adsabs.harvard.edu/abs/2024A&A...686A..84L}
}

@ARTICLE{Dufresne2024,
       author = {{Dufresne}, R.~P. and {Del Zanna}, G. and {Young}, P.~R. and {Dere}, K.~P. and {Deliporanidou}, E. and {Barnes}, W.~T. and {Landi}, E.},
        title = "{CHIANTI{\textemdash}An Atomic Database for Emission Lines{\textemdash}Paper. XVIII. Version 11, Advanced Ionization Equilibrium Models: Density and Charge Transfer Effects}",
      journal = {\apj},
         year = 2024,
        month = oct,
       volume = {974},
       number = {1},
          eid = {71},
        pages = {71},
          doi = {10.3847/1538-4357/ad6765},
archivePrefix = {arXiv},
       eprint = {2403.16922},
 primaryClass = {astro-ph.SR},
       adsurl = {https://ui.adsabs.harvard.edu/abs/2024ApJ...974...71D}
}

@ARTICLE{Dere1997,
       author = {{Dere}, K.~P. and {Landi}, E. and {Mason}, H.~E. and {Monsignori Fossi}, B.~C. and {Young}, P.~R.},
        title = "{CHIANTI - an atomic database for emission lines}",
      journal = {\aaps},
         year = 1997,
        month = oct,
       volume = {125},
        pages = {149-173},
          doi = {10.1051/aas:1997368},
       adsurl = {https://ui.adsabs.harvard.edu/abs/1997A&AS..125..149D}
}

@software{Dere2013,
       author = {{Dere}, Ken},
        title = "{ChiantiPy: Python package for the CHIANTI atomic database}",
 howpublished = {Astrophysics Source Code Library, record ascl:1308.017},
         year = 2013,
        month = aug,
          eid = {ascl:1308.017},
archivePrefix = {ascl},
       eprint = {1308.017},
       adsurl = {https://ui.adsabs.harvard.edu/abs/2013ascl.soft08017D}
}

@ARTICLE{Turner2021,
       author = {{Turner}, Jake D. and {Ridden-Harper}, Andrew and {Jayawardhana}, Ray},
        title = "{Decaying Orbit of the Hot Jupiter WASP-12b: Confirmation with TESS Observations}",
      journal = {\aj},
         year = 2021,
        month = feb,
       volume = {161},
       number = {2},
          eid = {72},
        pages = {72},
          doi = {10.3847/1538-3881/abd178},
archivePrefix = {arXiv},
       eprint = {2012.02211},
 primaryClass = {astro-ph.EP},
       adsurl = {https://ui.adsabs.harvard.edu/abs/2021AJ....161...72T}
}

@article{cox1987LocalInterstellarMedium,
  title = {The {{Local Interstellar Medium}}},
  author = {Cox, Donald P. and Reynolds, Ronald J.},
  year = {1987},
  month = sep,
  volume = {25},
  number = {Volume 25, 1987},
  pages = {303--344},
  publisher = {Annual Reviews},
  issn = {0066-4146, 1545-4282},
  doi = {10.1146/annurev.aa.25.090187.001511},
  urldate = {2025-05-29},
  language = {en},
  journal = {\araa}
}

@misc{draine2009InterstellarDustModels,
  title = {Interstellar {{Dust Models}} and {{Evolutionary Implications}}},
  author = {Draine, B. T.},
  year = {2009},
  month = dec,
  volume = {414},
  publisher = {arXiv},
  address = {eprint: arXiv:0903.1658},
  doi = {10.48550/arXiv.0903.1658},
  urldate = {2025-05-29}
}

@article{edenhofer2024ParsecscaleGalactic3D,
  title = {A Parsec-Scale {{Galactic 3D}} Dust Map out to 1.25 Kpc from the {{Sun}}},
  author = {Edenhofer, Gordian and Zucker, Catherine and Frank, Philipp and Saydjari, Andrew K. and Speagle, Joshua S. and Finkbeiner, Douglas and En{\ss}lin, Torsten A.},
  year = {2024},
  month = may,
  volume = {685},
  pages = {A82},
  issn = {0004-6361},
  doi = {10.1051/0004-6361/202347628},
  urldate = {2025-05-27},
  journal = {\aap}
}

@article{linsky2021CouldLocalCavity,
  title = {Could the {{Local Cavity}} Be an {{Irregularly Shaped Str{\"o}mgren Sphere}}?},
  author = {Linsky, Jeffrey L. and Redfield, Seth},
  year = {2021},
  month = oct,
  volume = {920},
  pages = {75},
  publisher = {IOP},
  issn = {0004-637X},
  doi = {10.3847/1538-4357/ac1feb},
  urldate = {2025-05-29},
  journal = {\apj}
}

@article{mccallum2025HaSkyThree,
  title = {The {{H$\alpha$}} Sky in Three Dimensions},
  author = {McCallum, Lewis and Wood, Kenneth and Benjamin, Robert A. and Krishnarao, Dhanesh and Zucker, Catherine and Edenhofer, Gordian and Haffner, L. Matthew},
  year = {2025},
  month = jun,
  volume = {540},
  pages = {L21-L27},
  publisher = {OUP},
  issn = {0035-8711},
  doi = {10.1093/mnrasl/slaf023},
  urldate = {2025-05-29},
  journal = {\mnras}
}

@article{oneill2024LocalBubbleLocal,
  title = {The {{Local Bubble Is}} a {{Local Chimney}}: {{A New Model}} from {{3D Dust Mapping}}},
  shorttitle = {The {{Local Bubble Is}} a {{Local Chimney}}},
  author = {O'Neill, Theo J. and Zucker, Catherine and Goodman, Alyssa A. and Edenhofer, Gordian},
  year = {2024},
  month = sep,
  volume = {973},
  number = {2},
  pages = {136},
  publisher = {The American Astronomical Society},
  issn = {0004-637X},
  doi = {10.3847/1538-4357/ad61de},
  urldate = {2025-05-29},
  language = {en},
  journal = {\apj}
}

@article{welsh2009TroubleLocalBubble,
  title = {The Trouble with the {{Local Bubble}}},
  author = {Welsh, Barry Y. and Shelton, Robin L.},
  year = {2009},
  month = sep,
  volume = {323},
  pages = {1--16},
  publisher = {Springer},
  issn = {0004-640X},
  doi = {10.1007/s10509-009-0053-3},
  urldate = {2025-05-29},
  journal = {\apss}
}

@article{zhang2023Parameters220Million,
  title = {Parameters of 220 Million Stars from {{Gaia BP}}/{{RP}} Spectra},
  author = {Zhang, Xiangyu and Green, Gregory M and Rix, Hans-Walter},
  year = {2023},
  month = sep,
  volume = {524},
  number = {2},
  pages = {1855--1884},
  issn = {0035-8711},
  doi = {10.1093/mnras/stad1941},
  urldate = {2025-05-29},
  journal = {\mnras}
}

@article{zucker2021ThreedimensionalStructureLocal,
  title = {On the {{Three-dimensional Structure}} of {{Local Molecular Clouds}}},
  author = {Zucker, Catherine and Goodman, Alyssa and Alves, Jo{\~a}o and Bialy, Shmuel and Koch, Eric W. and Speagle, Joshua S. and Foley, Michael M. and Finkbeiner, Douglas and Leike, Reimar and En{\ss}lin, Torsten and Peek, Joshua E. G. and Edenhofer, Gordian},
  year = {2021},
  month = sep,
  volume = {919},
  pages = {35},
  publisher = {IOP},
  issn = {0004-637X},
  doi = {10.3847/1538-4357/ac1f96},
  urldate = {2025-05-27},
  journal = {\apj}
}

\label{lastpage}

\end{document}